\documentclass[prd,nofootinbib,twocolumn]{revtex4}

\usepackage{amssymb}
\usepackage{amsmath} 
\usepackage{graphicx,subfigure} 
\usepackage{verbatim}
\usepackage{amsfonts}

\newcommand{\rep}[1]{(#1)}%
\newcommand{\ME}[2]{\left\langle #1|\rep{#2}|\bar{3} \right\rangle}%

\newcommand{\nn}{\nonumber}

\allowdisplaybreaks

\begin{document}

\begin{flushright}
DO-TH 13/24\\
TTP 14-027\\
\end{flushright}

\vspace*{-26mm}

\title{Standard Model Predictions and New Physics Sensitivity in \boldmath{$B\to DD$} Decays}

\author{Martin Jung}
\email{martin.jung@tum.de}
\affiliation{Institut f\"ur Physik, Technische Universit\"at Dortmund, 44221 Dortmund,~Germany\\ Excellence Cluster Universe, Technische Universit\"at M\"unchen, Boltzmannstr. 2, 85748 Garching, Germany}
\author{Stefan Schacht}
\email{stefan.schacht@kit.edu}
\affiliation{Institut f\"ur Theoretische Teilchenphysik, Karlsruher Institut f\"ur Technologie, D-76128 Karlsruhe, Germany}

\vspace*{1cm}

\begin{abstract}
An extensive model-independent analysis of $B\to DD$ decays is carried out employing SU(3) flavour symmetry, 
including symmetry-breaking corrections. Several theoretically clean observables are identified which allow for testing the Standard Model. These include the known time-dependent CP asymmetries, the penguin pollution of which can be controlled in this framework, but notably also quasi-isospin relations which are experimentally well accessible and unaffected by symmetry-breaking corrections. Theoretical assumptions can be kept to a minimum and controlled by additional sum rules.
Available data are used in global fits to predict the branching ratio for the $B^0\to D_s^+D_s^-$ decay as well as several CP asymmetries which have not been measured so far, and future prospects are analyzed.
\end{abstract}

\maketitle

\section{Introduction}
In the past decade, the connection of CP and flavour violation in the Standard Model (SM), embodied by the Kobayashi-Maskawa mechanism \cite{Kobayashi}, has been confirmed up to the order of $10-20\%$~\cite{CKMfitter,UTfit}. The fact that there is still no direct evidence of physics beyond the SM comes rather unexpected; as a consequence, we have to prepare to search for new physics (NP) effects which are small compared to the leading SM contributions, even when the latter are already suppressed. Specifically, in the context of quark flavour physics, this typically implies the necessity to gain (improved) control over hadronic matrix elements (MEs), which for most modes still constitutes a serious challenge. In this article, this is achieved by employing SU(3)-flavour symmetry including symmetry-breaking corrections, thereby extracting information on the MEs from data.
The full set of $B$-meson decays into two charmed pseudoscalars, $B\to DD$, is considered which constitutes a valuable source of information on the weak phases of the SM and provides access to NP.

Recently, the LHCb collaboration has measured various relative branching ratios~\cite{Aaij:2013fha,Aaij:2014pha} and two effective lifetimes~\cite{Aaij:2013bvd} of $B\to DD$ modes for the first time. %
Together with previous measurements \cite{Beringer:1900zz,Amhis:2012bh,Aubert:2006ia,Zupanc:2007pu,Aubert:2008ah,Rohrken:2012ta,Aaltonen:2012mg,Esen:2012yz} this enables 
a phenomenological analysis for the full set of modes which was not possible before. However, the flavour-SU(3) symmetry relating them is known to be broken at a level of $\epsilon_{\rm SU(3)}\sim m_s/\Lambda_{\rm QCD}\sim20-30\%$, making precision predictions difficult. Specifically, in Ref.~\cite{Gronau:2008ed} it has been demonstrated that the symmetry breaking severely affects the extraction of the CKM angle $\gamma$. Furthermore, also strategies to extract the weak mixing angles $\phi_{d,s}$ including ``penguin pollution'' by using (the U-spin subgroup of) SU(3) (as for example in Refs.~\cite{FleischerPsiK,Ciuchini2005,Fleischer:2007zn,FFJM}) are affected to some extent, as demonstrated recently in Refs.~\cite{Jung:2009pb,Jung:2012mp}.
This work includes therefore the SU(3)-breaking corrections model-independently. Since there are several suppression factors of similar order in $B\to DD$ decays, some SU(3)-breaking MEs are potentially larger than others appearing already in the SU(3) limit. We therefore develop a power counting to obtain information on both, the SU(3)-breaking and other subleading amplitudes like penguin and annihilation contributions from data, taking into account the full set of $B\to DD$ modes. While approximations remain necessary, they are made on a subleading level compared to previous analyses and can be tested within the resulting framework. Furthermore, the analysis will improve in the future due to the higher precision of the expected data, especially from the LHCb and Belle~II experiments~\cite{Alves:2008zz,Bediaga:1443882,Abe:2010gxa}.

This article is organized as follows: Sec.~\ref{sec::SU3} provides the SU(3) analysis of the modes under consideration, including SU(3) breaking model-independently. The following definition of the power counting allows to identify three quasi-isospin relations. The section closes with a discussion of parametrization invariance (RI), which crucially affects the extraction of the angle~$\gamma$. 
The phenomenological analysis is carried out in Sec.~\ref{sec::pheno}, where the presently available data are analyzed, sum rules for amplitudes and rates derived and 
key observables identified to test the SM in the future. The following global analysis allows for predicting various CP asymmetries and one rate in $B\to DD$ that have not been measured yet. 
In Sec.~\ref{sec::NP} explicit inclusion of NP contributions in the analysis is discussed, before concluding in Sec.~\ref{sec::conclusions}; several Appendices contain additional details on the experimental input, the SU(3) analysis, the performed fits, and the uncertainty estimates for future measurements.

\section{\boldmath $B\to DD$ amplitudes in broken SU(3)\label{sec::SU3}}
In this section the theoretical analysis of the $B\to DD$ amplitudes is performed. First the necessary SU(3) expressions including symmetry breaking are derived, then the  power counting for the various suppression effects defined and from that important model-independent amplitude relations derived. Furthermore, the relation of parametrization invariance and SU(3) breaking is discussed, questioning the sensitivity to the CKM angle $\gamma$ of these modes.

\subsection{SU(3) limit\label{sec::SU3limit}}
We start by performing the SU(3) analysis in the symmetry limit. Ignoring electroweak penguin operators with very small Wilson coefficients (to which we will return below), the relevant effective hamiltonian reads
\begin{eqnarray}
\mathcal{H}_{\rm eff}^{b\to d,s} &=& \frac{4G_F}{\sqrt{2}}\!\sum_{U=c,u}\sum_{D=d,s}\!\!\lambda_{UD}\!\left\{\sum_{i=1}^2C_i\mathcal{O}_i^U+\sum_{i=3}^6 C_i\mathcal{O}_i\right\}\nonumber\\
&\equiv&\mathcal{H}_c+\mathcal{H}_u\,,
\end{eqnarray}
with $\lambda_{UD}=V_{Ub}V_{UD}^*$, the tree operators $\mathcal{O}_{1,2}^{u,c}$ and the penguin operators $\mathcal{O}_{3-6}$, see e.g. Ref.~\cite{BBL}. The symmetry analysis is analogous to the one in Ref.~\cite{Jung:2012mp} (see also Refs.~\cite{Zeppenfeld,Savage:1989ub} for early applications), leaving us with a pure flavour triplet coming with $\lambda_{cD}$, while the tree operators $\mathcal{O}_{1,2}^u$ involve the  representations $R=3,\bar 6,15$. With the initial state~$i$ transforming as a $\bar3$ and the final state~$f$ as $1\oplus 8$, this implies a description with six independent reduced MEs. The decay amplitude of a $b\to D$ decay $\mathcal{D}$ is thus expressed as 
\begin{eqnarray}
\mathcal A_\mathcal{D}&=&\mathcal A_c(\mathcal{D})+\mathcal A_u(\mathcal{D})\nonumber\\
&=& \sum_{U=u,c}\lambda_{UD}\sum_{R,f}C_{f\,R}^U(\mathcal{D})\langle f|R|i\rangle_U\,,\label{eq::CfRUdef}
\end{eqnarray} 
with the Clebsch-Gordan coefficients $C_{f\,R}^U(\mathcal{D})$ provided in Table~\ref{tab::SU3coeffs} in Appendix~\ref{app::details}, in accordance with Refs.~\cite{Zeppenfeld,Savage:1989ub}.

\begin{table}
\begin{tabular}{l c c c c c c}
\hline\hline
Mode						& $\lambda_{cD}T$	& $\lambda_{cD}A^c$	& $\lambda_{uD}\tilde P_1$	& $\lambda_{uD}\tilde P_3$	& $\lambda_{uD}A^u_1$	& $\lambda_{uD}A^u_2$\\\hline\hline
Counting					& 1					& $\delta^2$		& $\delta^{3(5)}$ 			& $\delta^{4(6)}$			& $\delta^{3(5)}$		& $\delta^{4(6)}$\\\hline\hline
$B^-\to D^- D^0$			& 1					& 0					& -1						& 0							& 1						& 0\\
$B^-\to D^-_s D^0$			& 1					& 0					& -1						& 0							& 1						& 0 \\
$\bar B^0\to D_s^-D^+$		& 1					& 0					& -1						& 0							& 0						& 0 \\
$\bar B_s\to D^-D_s^+$		& 1					& 0					& -1						& 0							& 0						& 0 \\
$\bar B^0\to D^-D^+$		& 1					& 1					& -1						& -1						& 0						& 0 \\
$\bar B_s\to D_s^-D_s^+$	& 1					& 1					& -1						& -1						& 0						& 0 \\
$\bar B^0\to D_s^-D_s^+$	& 0					& 1					& 0							& -1						& 0						& 0 \\
$\bar B_s\to D^-D^+$		& 0					& 1					& 0							& -1						& 0						& 0 \\
$\bar B^0\to \bar D^0D^0$	& 0					& -1				& 0							& 1							& 0						& -1 \\
$\bar B_s\to \bar D^0D^0$	& 0					& -1				& 0							& 1							& 0						& -1
\\\hline\hline
\end{tabular}
\caption{\label{tab::topamps}$B\to DD$ amplitudes in the SU(3) limit, given in terms of topological amplitudes, see Appendix~\ref{app::details} for details. The power counting is explained in the text; the number in brackets indicates the additional CKM suppression in $b\to s$ transitions.}
\end{table}

Approximations typically involve arguments about different contractions of the involved operators which are commonly described in terms of topological amplitudes \cite{Chau:1982da,Gronau:1994rj,Gronauetal1,BurasSilvestrini}, which can, however, be expressed in terms of SU(3) MEs \cite{Zeppenfeld}. We therefore translate the derived SU(3) amplitudes into this language, using the topological amplitudes introduced in Ref.~\cite{BurasSilvestrini}. Note that the number of topologies before redefinitions is larger than the number of SU(3) amplitudes. The description is, however, equivalent, as long as no assumptions regarding the  various MEs are made.
The translation is again given in Appendix~\ref{app::details}, the resulting amplitude decompositions in the SU(3) limit are given in Table~\ref{tab::topamps}.

\subsection{Including SU(3) breaking}
SU(3) breaking is induced by the quark mass term, transforming as an octet when neglecting isospin breaking \cite{Savage:1991wu} (see also Refs.~\cite{Gronauetal1,GrinsteinLebed} for this treatment of symmetry breaking in $B$ decays). The leading SU(3)-breaking part of $\mathcal A_c$ consists of the four MEs of its tensor product with the effective hamiltonian; the corresponding coefficients are listed in Table~\ref{tab::SU3coeffseps} in Appendix~\ref{app::details}, and in Table~\ref{tab::topampseps} again translated to topological amplitudes. After redefinitions, there are furthermore five SU(3)-breaking MEs at first order in $\mathcal A_u$; as discussed below, however, for all considered observables they have only a small impact  compared to the present experimental precision. For the sake of completeness, these coefficients are presented in Appendix~\ref{app::details}, but we will only use the corrections to $\mathcal A_c$ in the phenomenological analysis.

\begin{table}
\begin{tabular}{l c c c c}
\hline\hline
Mode						& $\lambda_{cD}\delta T_1$	& $\lambda_{cD}\delta T_2$	& $\lambda_{cD}\delta A^c_{1}$	& $\lambda_{cD}\delta A^c_{2}$	\\\hline\hline
Counting					& $\delta$					& $\delta$					& $\delta^{3}$ 					& $\delta^{3}$	\\\hline\hline
$B^-\to D^- D^0$			& 0							& $-\frac{1}{2}$			& 0								& 0\\
$B^-\to D^-_s D^0$			& 1							& 0							& 0								& 0\\
$\bar B^0\to D_s^-D^+$		& 1							& 0							& 0								& 0\\
$\bar B_s\to D^-D_s^+$		& -1						& $\frac{1}{2}$				& 0								& 0\\
$\bar B^0\to D^-D^+$		& 0							& $-\frac{1}{2}$			& $\frac{1}{2}$					& $-\frac{1}{2}$\\
$\bar B_s\to D_s^-D_s^+$	& 0							& 1							& -1							& 1\\
$\bar B^0\to D_s^-D_s^+$	& 0							& 0							& $\frac{1}{2}$					& $\frac{1}{2}$\\
$\bar B_s\to D^-D^+$		& 0							& 0							& -1							& 0\\
$\bar B^0\to \bar D^0D^0$	& 0							& 0							& $-\frac{1}{2}$				& $\frac{1}{2}$\\
$\bar B_s\to \bar D^0D^0$	& 0							& 0							& 1								& 0							
\\\hline\hline
\end{tabular}
\caption{\label{tab::topampseps}First order SU(3)-breaking corrections to the $\mathcal A_c$ part of the $B\to DD$ amplitudes, given in terms of topological amplitudes, see Appendix~\ref{app::details} for details. The power counting is explained in the text.}
\end{table}

Extending this analysis to higher orders in the symmetry breaking yields at most one additional ME in both amplitudes $\mathcal A_{u,c}$: this observation follows from the analysis in Ref.~\cite{GrinsteinLebed}, where all potentially contributing MEs are listed (without their hierarchy). The expressions given here are therefore already close to the most general result. The missing pieces, which can be counted as $\mathcal{O}(\epsilon^2)$, can be  derived from that result and are given in Table~\ref{tab::SU3coeffseps2} again for completeness.

\subsection{Power counting\label{sec::pc}}
As a prerequisite for the phenomenological discussion, the expected size of the various contributions has to be classified. This is non-trivial for several reasons, one issue being that there are various suppression factors involved:
\begin{itemize}
\item \textbf{CKM structure} $|\lambda_{uD}/\lambda_{cD}|=\{R_u,\lambda^2R_u\}$ for $D=\{d,s\}$, respectively, where $R_u\approx 0.35$ denotes a side in the unitarity triangle, implying in both cases a suppression of $\mathcal A_u(\mathcal{D})$ relative to $\mathcal A_c(\mathcal{D})$, but especially for $D=s$.
\item \textbf{Penguin suppression} The amplitudes $\tilde P_i$ have two contributions: MEs of penguin operators, which are suppressed by the small Wilson coefficients $C_{3-6}\sim {\rm few}\,\%$, and penguin contractions of the tree operators $\mathcal{O}_{1,2}^u$, which involve additional interactions to create the $\bar cc$ pair, yielding a similar suppression which is however harder to quantify.
\item \textbf{Annihilation} The annihilation graphs $A_i^U$ involve the spectator quark, implying naively a suppression of the order $\Lambda_{\rm QCD}/m_b$. However, non-factorizable contributions can give larger contributions, \emph{e.g.} of order $m_D/m_B$ \cite{Eeg:2003yq}. Nevertheless, these contributions remain suppressed.
An additional suppression is assumed when the interaction furthermore involves the creation of a $\bar cc$ pair from the vacuum, which is the case for the amplitudes $A_i^u$ \cite{Gronau:2008ed}.
\item \textbf{$\mathbf{1/N_c}$ suppression} The various topologies can be classified according to their scaling with the number of colours $N_c=3$ \cite{'tHooft:1973jz,Buras:1985xv}. The scaling for the amplitudes in Table~\ref{tab::topamps} reads \cite{BurasSilvestrini} $T,A^u_1\sim1$, $\tilde P_1,A_2^u,A^c\sim1/N_c$ and $\tilde P_3\sim1/N_c^2$.
\end{itemize}
To obtain a power counting for the amplitudes, we assign to (each order of) these effects as well as for SU(3) breaking a common factor $\delta\sim20-30\%$, yielding $T\sim 1, A^c\sim\delta^2, \lambda_{uD}\tilde P_1,\lambda_{uD}A_1^u\sim\delta^{3(5)}$ and $\lambda_{uD}A_2^u,\lambda_{uD}\tilde P_3\sim\delta^{4(6)}$ for $D=d(s)$, with an additional factor of $\delta$ for the SU(3) corrections to these amplitudes. Note that these factors are given relative to the leading amplitude, \emph{i.e.} relative to $\lambda_{cD}$ for a $b\to D$ decay. While this  assignment is not rigorous, it allows for a systematic classification of the amplitudes in question. Despite some of these arguments being on the level of topological amplitudes which can rescatter into each other, the estimates are expected to be conservative enough to include these effects. 
Below we will differentiate between predictions expected to hold generally in the SM, like the quasi-isospin relations, and others more sensitive to dynamical assumptions; we furthermore identify experimental tests for both types.

A first important consequence of our power counting 
is that all amplitudes $\mathcal A_u(\mathcal{D})$ are suppressed at least like $\delta^2$ relative to the leading amplitude $\mathcal A_c(\mathcal{D})$. 
In the following we will neglect therefore the SU(3) corrections to the former, 
given the present experimental precision for the CP asymmetries; 
this assumption should be reconsidered in the future, but then measurements will signal this necessity and allow for an improved fit in any case. Using this approximation, 10 unknown MEs remain. 
While in principle these modes offer up to 26 observables, with the limited data available 
the power counting is necessary to make a fit viable.
To determine the influence of different assumptions and inputs, three scenarios are introduced in Sec.~\ref{sec::pheno}, yielding predictive frameworks and at the same time testing the power counting.

\subsection{Quasi-isospin relations\label{sec::isospin}}
In absence of electroweak penguin operators, the $b\to s$ part in $\mathcal{H}_c$ transforms as a pure isospin singlet~\cite{Lipkin:1987xw}. This results in the relations
\begin{eqnarray}
\mathcal A_c(\bar B^0\to D_s^-D^+) &=& \phantom{-}\mathcal A_c(B^-\to D_s^-D^0)\,\,\,{\rm and}\label{eq::isorel1}\\
\mathcal A_c(\bar B_s\to D^-D^+) &=& -\mathcal A_c(\bar B_s\to \bar D^0D^0)\,,\label{eq::isorel2}
\end{eqnarray}
which do not receive SU(3)-breaking corrections, see also Ref.~\cite{Gronauetal1}. Importantly, the same relations hold for the penguin contributions to $\mathcal A_u$: extending them to the full amplitudes, the only corrections stem from highly suppressed annihilation contributions in $\mathcal A_u$ ($\lambda_{us}A_1^u$ and $\lambda_{us}A_2^u$ at $\mathcal{O}(\delta^{5(6)})$ in our power counting for the first (second) relation) and $\Delta I=1$ contributions from electroweak penguin operators in $\mathcal A_c$, which are heavily suppressed as well. Therefore, we have actually for the \emph{full} amplitudes 
\begin{eqnarray}
\mathcal A_{\bar B^0\to D_s^-D^+} &\simeq& \phantom{-}\mathcal A_{B^-\to D_s^-D^0}\quad{\rm and}\label{eq::isorel1full}\\
\mathcal A_{\bar B_s\to D^-D^+} &\simeq& -\mathcal A_{\bar B_s\to \bar D^0D^0}\,,\label{eq::isorel2full}
\end{eqnarray}
to very high precision in the SM, even in the presence of enhanced penguin contributions and still unaffected by SU(3) breaking.
Similarly, for $b\to d$ decays the following quasi-isospin relation arises:
\begin{equation}
\mathcal A_{\bar B^0\to D^-D^+}+\mathcal A_{\bar B^0\to \bar D^0D^0} \simeq \mathcal A_{B^-\to D^-D^0}\,,\label{eq::isorel3}
\end{equation}
which again receives corrections from annihilation contributions, this time at the level of $\mathcal{O}(\delta^{3})$, and $\Delta I=3/2$ contributions from electroweak penguin operators; this last relation has also been discussed in Refs.~\cite{Sanda:1996pm,Gronau:2008ed}. Note that $|\mathcal A_{\bar B^0\to \bar D^0D^0}|$ is much smaller than the other two amplitudes. 
Note furthermore that all three quasi-isospin relations hold analogously for $B\to DD^*$ and $B\to D^*D^*$ modes.

Given the very high precision of especially Eqs.~\eqref{eq::isorel1full} and~\eqref{eq::isorel2full},
we would like to comment on the potential influence of electroweak penguin operators. Their Wilson coefficients are very small, which we count conservatively as $\mathcal O(\delta^2)$. But most importantly, their main contributions stem from the SU(3)-triplet part (including the operators with the flavour structure $(\bar D b)(\bar c c)$), which have tree-level MEs with the final state; these contributions can be absorbed into the MEs already present. The remaining contributions, which involve a change in isospin by $\Delta I=1,3/2$ (and therefore lead to corrections to Eqs.~\eqref{eq::isorel1}-\eqref{eq::isorel3}), stem from insertions of the electroweak penguin operators into annihilation diagrams, which then require additionally the creation of a $\bar cc$ pair from the vacuum and are suppressed by $N_c$, yielding contributions suppressed at least as  $\mathcal{O}(\delta^5)$ in $\mathcal A_c(\mathcal{D})$ and $\delta^6$ in $\mathcal A_u(\mathcal{D})$. We neglect these tiny terms in the following and consider the leading ones absorbed into the MEs already present. 

In the SM, Eqs.~\eqref{eq::isorel1full} and~\eqref{eq::isorel2full} therefore directly translate into precision relations for the corresponding rates, unaffected by the $SU(3)$ symmetry breaking, thereby allowing to test for NP with $\Delta I=1$. Note that the corresponding ratios of rates are also experimentally advantageous, since they are independent of $f_s/f_d$. Similarly, Eq.~\eqref{eq::isorel3} provides a test for $\Delta I=3/2$ contributions, albeit in this case with a SM ``pollution'' of a few per cent. The corresponding CP asymmetries are predicted to be small in all $b\to s$ modes; any sizable signal (more than $\sim 3(10)$ per cent for tree-(annihilation-)dominated modes and enhanced penguins) would imply NP as well. Furthermore, the potentially large \emph{relative} corrections for the CP asymmetries provide access to the amplitudes $A_i^u$. 
We discuss additional sum rules for amplitudes and observables in Sections~\ref{sec::sumrulesa} and~\ref{sec::sumrulesr}.

\subsection{Reparametrization invariance \label{sec::RI}}
Before turning to the phenomenological analysis, we would like to comment on an effect prohibiting in some situations the extraction of the weak phase $\gamma$, usually dubbed ``reparametrization invariance'' (RI) \cite{London:1999iv,Botella:2005ks,Feldmann:2008fb}: the basic observation is that, depending on the structure of the decays in question, there exist transformations that leave the decay amplitudes form-invariant, but change the apparent weak phase. As a result, it is impossible in these situations to extract $\gamma$ without knowledge of the MEs; only the values $0$ and $\pi$ can be excluded from data. This situation usually changes when considering more decay modes which are related by flavour symmetries, since the different MEs enter with different weight, breaking the invariance. This is confirmed for $B\to DD$ decays by observing that the coefficient matrix given in Table~\ref{tab::topamps} has full rank when including the CKM factors.

But the situation is actually more subtle. First of all, considering \emph{e.g.} only $b\to d$ modes, which are expected to have larger direct CP asymmetries and thereby increased sensitivity to $\gamma$, we still observe an approximate RI, broken only by the MEs $A_{1,2}^u$. The same statement holds when considering only $b\to s$ modes. Again, when combining the two sets, the RI is broken, even when neglecting $A_{1,2}^u$. However, we observe that the inclusion of unknown $SU(3)$-breaking MEs can actually  \emph{restore} the approximate RI: the additional MEs are capable of absorbing the RI-breaking terms, implying again no sensitivity to $\gamma$ as long as $A_{1,2}^u$ can be neglected. This is the analytical reason behind the observation of very large uncertainties for $\gamma$ in the presence of general $SU(3)$ breaking in Ref.~\cite{Gronau:2008ed}. Related to the RI is a reduced rank of the coefficient matrix including the CKM factors: the combinations of MEs entering are such that the different relative weights from $\lambda_{ud},\lambda_{us}$ do not enter explicitly, allowing to perform the redefinition for the full set of amplitudes.

It should be emphasized that the (assumed) knowledge of one or several of the MEs will break the invariance, as will to some extent restrictions on the size of MEs. For example, the assumption of factorizable $SU(3)$ breaking in Refs.~\cite{FleischerPsiK,Datta:2003va} enables the extraction of $\gamma$ from $B\to D^+D^-$ and $B_s\to D_s^+D_s^-$ ($B^0\to D_s^+D^-$) in these cases. However, the theoretical error related to this assumption is hard to quantify, prohibiting a precision extraction. Also applying our power counting restricts the freedom to redefine the various parameters, but again high precision seems unachievable unless there is substantial theoretical progress in calculating the relevant hadronic amplitudes explicitly.

We will therefore concentrate on the mixing phase $\phi_{s}$, the extraction of which remains mostly unaffected by these observations, and use external input for $\phi_d$ and $\gamma$ which breaks the invariance and allows for extracting hadronic parameters.
Note that using the SM central value for $\gamma$ is no restriction: due to the RI, also NP amplitudes can be written with that phase; only the interpretation of the hadronic parameters changes in this case.

\section{Phenomenological Analysis\label{sec::pheno}}

After a review of the experimental situation for $B\to DD$ decays we define three scenarios to analyze the relative importance of the various dynamical suppression effects. Next we derive for each of them sum rules for amplitudes and observables. 
We discuss 
the data for branching ratios and CP asymmetries with respect to the amplitude structure obtained in the last section,
before discussing the results of the global analysis and its future prospects.

\subsection{Experimental situation\label{sec::exp}}
Until recently, only little data were available for $B\to DD$ decays. This situation has improved, starting by first measurements of $B_s$ decays by the CDF \cite{Abulencia:2007zz,Aaltonen:2012mg} and Belle \cite{Esen:2010jq,Esen:2012yz} experiments, and more recently by results from the LHCb experiment \cite{Aaij:2013fha,Aaij:2013bvd,Aaij:2014pha}. Specifically, Ref.~\cite{Aaij:2013fha} provides relative branching ratios for all $B_s$ modes and several $B^0$ modes as well, leaving only the rate for $B^0\to D_s^+D_s^-$ undetermined. In addition, several measurements of CP asymmetries are expected from LHCb in the near future, given the much larger data sample that has become available since. In the farther future, also the Belle~II experiment is expected to deliver data for these modes. These experimental developments are a main motivation of the present study. 

In Table~\ref{tab::data}, we list the presently available data for all $B\to DD$ modes. In the cases where relative branching ratios have been measured, we use them in the numerical analysis, providing here the absolute ones only for convenience. For $B^-\to D_s^-D^0$, $\bar B^0\to D^+D^-$ and $\bar B_s\to D_s^+D_s^-$, we combine the very recent measurements by the LHCb \cite{Aaij:2013fha} and Belle \cite{Rohrken:2012ta,Esen:2012yz} collaborations with those considered already in the latest PDG update~\cite{Beringer:1900zz}. Furthermore we include a correction stemming from $\Gamma(\Upsilon(4S)\to B^+B^-)/\Gamma(\Upsilon(4S)\to \bar B^0 B^0)=1.055\pm0.025$ \cite{Beringer:1900zz} for the $B$ factory results for $\bar B^0\to D^+D^-$ and $B^-\to D^0D^-$.

For a decay $\mathcal D$ of a $\bar B$ meson (at $t=0$) into a CP eigenstate $f$, we use the following notation for the time-dependent CP asymmetries:
\begin{eqnarray}\label{eq::DefCPA}
a_{\rm CP}(\mathcal D;t)\! &\equiv&\! \frac{\Gamma(\mathcal D;t)-\Gamma(\bar{\mathcal D};t)}{\Gamma(\mathcal D;t)+\Gamma(\bar{\mathcal D};t)}\nonumber\\
&=&\! \frac{S_{\rm CP}(\mathcal D)\sin(\Delta m t)+A_{\rm CP}(\mathcal D)\cos(\Delta m t)}{\cosh(\Delta \Gamma t/2)-A_{\Delta \Gamma}(\mathcal D)\sinh(\Delta \Gamma t/2)}.\,
\end{eqnarray}
The coefficients for mixing-induced CP violation $S_{\rm CP}(\mathcal D)$, direct CP violation $A_{\rm CP}(\mathcal D)$ and the rate asymmetry $A_{\Delta\Gamma}(\mathcal D)$ are defined in Appendix~\ref{sec::AppExp}.
\begin{table}
\begin{tabular}{l c c c}\hline\hline
Mode						& $BR_{\rm theo}/10^{-3}$							& $A_{\rm CP}/\%$					& $S_{\rm CP}/\%$\\\hline\hline
$B^-\to D^- D^0$			& $0.37\pm0.04^\S$									& $3\pm7$ \cite{Beringer:1900zz}	& ---\\
$B^-\to D^-_s D^0$			& $9.4\pm0.9$										& 									& ---\\
$\bar B^0\to D_s^-D^+$		& $7.6\pm0.7$ 										& $-1\pm2^*$ \cite{Rohrken:2012ta} & ---\\
$\bar B_s\to D^-D_s^+$		& $0.30\pm0.04$										&									& ---\\
$\bar B^0\to D^-D^+$		& $0.226\pm0.023^\S$								& \textbf{(a)} $31\pm14$ \cite{Amhis:2012bh} & $-98\pm17$ \cite{Amhis:2012bh}\\
							& BaBar \cite{Aubert:2008ah}: 						& \textbf{(b)} $7\pm23$ 							& $-63\pm36$ \\
							& Belle \cite{Rohrken:2012ta}:						& $43\pm17$							& $-106^{+22}_{-16}$\\
$\bar B_s\to D_s^-D_s^+$	& $4.6\pm0.5$ 										& 						& \\
$\bar B^0\to D_s^-D_s^+$	& $\leq 0.036^\dagger$ \cite{Zupanc:2007pu}			& 									& \\
$\bar B_s\to D^-D^+$		& $0.27\pm0.05$ 									&									& \\
$\bar B^0\to \bar D^0D^0$	& $0.013\pm0.006$								 	&									& \\
$\bar B_s\to \bar D^0D^0$	& $0.19\pm0.04$ 									&									& 
\\\hline\hline
\end{tabular}
\caption{\label{tab::data} Available data for $B\to DD$ decays. The values for the branching ratios correspond to a fit to the data in Table~\ref{tab::adddata} (statistical and systematic uncertainties added in quadrature), taking the finite width of the $B_s$ into account, see text. The labels (a) and (b) indicate the different inputs for the two datasets.\newline ${}^\S$:~Correction for $\Gamma(\Upsilon(4S)\to B^0\bar B^0)\neq \Gamma(\Upsilon(4S)\to B^+B^-)$ included. ${}^*$:~Statistical uncertainty only, not used in the numerical analysis. ${}^\dagger$:~Upper limit at $90\%$~CL, not used in the numerical analysis.}
\end{table}
An important issue is the difference between the results from the BaBar \cite{Aubert:2008ah} and Belle \cite{Rohrken:2012ta} experiments for the time-dependent CP asymmetry in $B^0\to D^+D^-$. The BaBar result lies within the physical region and is consistent with the expectations $|\Delta S_{\rm CP}(B^0\to D^+D^-)|,|A_{\rm CP}|\ll 1$ discussed below, where 
\begin{equation}
\Delta S_{\rm CP}(\mathcal{D})\equiv -\eta_{\rm CP}^fS_{\rm CP}(\mathcal{D})-\sin\phi(\mathcal{D})
\end{equation}
is the ``penguin shift''
and $\phi(\mathcal{D})=\phi_d(\phi_s)$ for a $B^0$ ($B_s$) decay. The Belle result, on the other hand, is outside the physical region (defined by $S_{\rm CP}^2+A_{\rm CP}^2\leq 1$), indicating large values for the direct CP asymmetry as well as $\Delta S_{\rm CP}(B^0\to D^+D^-)$. This makes the averaging of the two results problematic. The two datasets defined below are used to demonstrate the influence of these differences.

For modes with a low rate, it might be experimentally advantageous to start by measuring the time-integrated CP asymmetry.
However, this is only feasible for the $B_d$ modes, since the oscillations in the $B_s$ system are so fast that the asymmetry is mostly averaged out. 

While the values for the branching ratios quoted in Table~\ref{tab::adddata} correspond to the time-integrated rates, the  theoretical expressions for the rates below are for $t=0$. This difference is relevant only for $B_s$ decays into CP eigenstates; it is included following Ref.~\cite{DeBruyn:2012wj}, using 
\begin{equation}
BR_{\rm theo}=\frac{1-y_s^2}{1-A_{\Delta \Gamma}(\mathcal D)\,y_s}BR_{\rm exp}\,,
\end{equation}
where the value employed for the relative width difference is $y_s\equiv\Delta\Gamma_s/(2\Gamma_s)=0.067\pm0.008$ \cite{Amhis:2012bh}. The relative rate difference $A_{\Delta \Gamma}$ of heavy and light $B_s$ eigenstates to a given final state vanishes for flavour specific decays, and is generally final state dependent. It is related to the CP asymmetries by $(A_{\Delta \Gamma}(\mathcal D))^2+(S_{\rm CP}(\mathcal D))^2+(A_{\rm CP}(\mathcal D))^2=1$; we use for the calculation the theoretical expectation 
$A_{\Delta \Gamma}^{b\to s}(\mathcal D)\approx\cos\phi_s\approx1$, with a positive sign determined by the positive CP eigenvalue of the relevant final states, implying a rather large correction of $\sim 7\%$ for the corresponding branching ratios.
This result follows already from the observation that the relevant decays, \emph{i.e.} $B_s$ decays into CP eigenstates, are all $b\to s$ transitions: the CKM suppression alone for $A_{\rm CP}$ and $\Delta S_{\rm CP}$, together with $|\phi_s|\sim\mathcal O(\%)$ suffices to predict $|A_{\Delta\Gamma}^{b\to s}|\simeq\cos\phi_s=1+\mathcal O(10^{-3})$. This is even strengthened by dynamical considerations, leading to $1-|A_{\Delta\Gamma}^{b\to s}|\lesssim 10^{-3}$, implying excellent null tests of the SM from these observables. From these considerations the prediction for the effective lifetimes of these modes reads
\begin{equation}
\tau_{\rm eff}(\bar B_s\to DD|_{CP})=(1.421\pm0.013)~{\rm ps},
\end{equation}
using $\tau_s=(1.516\pm0.008)\,{\rm ps}$~\cite{Amhis:2012bh}, in agreement with the experimental result in Ref.~\cite{Aaij:2013bvd}, $\tau_{\rm eff}(\bar B_s\to D_s^-D_s^+)=(1.379\pm 0.026\pm 0.017)~{\rm ps}$. 

Finally, we point out the importance of correlations between the measurements for $B_s$ decays, especially those induced by the ratio of production rates $f_s/f_d$. This ratio provides already the dominant systematic uncertainty for the ratios involving $B_s$ decays in Ref.~\cite{Aaij:2013fha} (except for $\bar B_s\to\bar D^0D^0$) and is likely to dominate soon the total uncertainty in various decays. At least in the latter situation the inclusion of the resulting correlations is mandatory and provides in addition the possibility to profit from future potentially more precise determinations of this quantity. For the LHCb results we therefore use this ratio explicitly in the numerical analysis, with its experimental value 
$f_s/f_d|_{\rm LHCb}=0.259\pm0.015$ \cite{Aaij:2013qqa,LHCb-CONF-2013-011}; for the CDF and Belle measurements this is not necessary, since the corresponding ratio, different from the one at LHCb, appears only once.\footnote{Note, however, that we use for the computation of the Tevatron value for $BR(\bar B_s \to D_s^-D_s^+)$ the ratio $f_s/f_d|_{\rm Tev}=0.311\pm0.037$ \cite{Amhis:2012bh} instead of the average of the Tevatron and LEP results used in Ref.~\cite{Aaltonen:2012mg}.}

\subsection{Scenarios\label{sec::scenarios}}

Given the presently available data, the necessity for approximations on a subleading level remains. To this aim again the power counting is applied. However, this counting may be spoiled when the schematic scaling with $\delta$ is violated by one or more of the discussed effects. 
Therefore, we define 
three scenarios, corresponding to different dynamical situations, which will help to judge the effectiveness of the various suppression mechanisms: 
\begin{enumerate}
\item[\textbf{S1:}] \textbf{SU(3) limit} We start by assuming the SU(3) hierarchy to be the strongest one and neglecting all SU(3)-breaking contributions;
given the discussion above, we do not expect a very good description of the data.
The  advantage of this limit is that we can test for the necessity of SU(3) breaking, and potentially also for subleading amplitudes in $\mathcal A_u$. 
\item[\textbf{S2:}] \textbf{Standard counting} Here the assumption is for all subleading MEs
to obey the power counting. This corresponds to the situation generically expected in the SM, already allowing for rather conservative ranges for the subleading contributions. The first contribution to $\mathcal A_u$ arises with the standard counting (S2))on the $\delta^{3(4)}$ level for the tree-dominated (annihilation) modes, implying typical direct CP asymmetries of a few per cent ($10\%$) for $b\to d$ decays and even less for $b\to s$ modes,\footnote{Note that in the fit we parametrize all MEs relative to $T$. Since we do not impose $|A^u_2/A^c|,|\tilde P_3/A^c|\sim\mathcal O(\delta^{(3)})$ additionally, larger values for $A_{\rm CP}$ might be allowed in the fit when $|A^c|$ is smaller than its expectation.} together with
$\Gamma(\mathcal D)\approx |\mathcal A_c(\mathcal D)|^2$.

The data for the CP asymmetries in $\bar B^0\to D^+D^-$ contradict the assumption in this scenario in part, as the Belle measurement indicates a sizable direct CP asymmetry as well as a deviation from $\Delta S_{\rm CP}(B\to D^+D^-)\approx 0$. This second scenario therefore corresponds to resolving the tension between the BaBar and Belle measurements in favour of the former. 
\item[\textbf{S3:}] \textbf{Enhanced penguins} The final scenario is defined by assuming no additional suppression for penguin MEs (as motivated by the largish CP asymmetries measured by Belle), such that $\mathcal A_u$ arises on the $\delta^2$ level. The ranges outside of S2, but inside the ones from this scenario can be interpreted as either extremely conservative SM predictions or already as NP.
\end{enumerate}
These scenarios allow for distinguishing different dynamical suppression effects. Significant measurements of observables outside the predictions of scenario~3 can be considered a NP signal.
The technical implementation of the power counting is that each real parameter $x$ related to a subleading ME of order $n$ is restricted to $|x|\leq \delta^{n-1/2}$, choosing $\delta=30\%$.

\subsection{Amplitude sum rules\label{sec::sumrulesa}}
The scenarios defined in the previous subsection imply relations between the amplitudes which hold up to SU(3)-breaking terms in S1 and up to corrections of higher orders in $\delta$ for scenarios~2 and~3. These can be used to further test the underlying assumptions. The number of sum rules is determined by the rank of the full coefficient matrix of the hadronic amplitudes, including the CKM factors; in general there exist sum rules even if the number of hadronic MEs equals the number of decays.
This is illustrated by the fact that there are two sum rules that hold exactly in all scenarios discussed here, reading 
\begin{eqnarray}
&&\mathcal A_{\bar B^0\to D_s^-D^+}-\mathcal A_{B^-\to D_s^-D^0}+\mathcal A_{\bar B_s\to D^-D^+}+\mathcal A_{\bar B_s\to \bar D^0D^0}\nonumber\\
&&=\frac{\lambda_{us}}{\lambda_{ud}}\left(\mathcal A_{\bar B^0\to D^-D^+}-\mathcal A_{B^-\to D^-D^0}+\mathcal A_{\bar B^0\to \bar D^0 D^0}\right)\,,\label{eq::SRfull1}\\
&&\mathcal A_{\bar B^0\to D_s^-D^+}-\mathcal A_{\bar B_s\to D_s^-D_s^+}+\mathcal A_{\bar B_s\to D^-D^+}=\nonumber\\
&&\frac{\lambda_{cs}}{\lambda_{cd}}\left(\mathcal A_{\bar B^0\to D^-D^+}-\mathcal A_{\bar B_s\to D^-D_s^+}-\mathcal A_{\bar B^0\to D_s^-D_s^+}\right)\,.\label{eq::SRfull2}
\end{eqnarray}
The first of these equations corresponds to a linear combination of the three quasi-isospin sum rules, Eqs.~\eqref{eq::isorel1full}-\eqref{eq::isorel3}, in which the very small terms in $\mathcal A_u$ breaking those relations cancel as well. It is only broken by an SU(3)-breaking correction to the annihilation amplitudes in $\mathcal A_u$. In the latter relation, an analogous cancellation happens for the more generic SU(3)-breaking contributions, including even the first-order ones  to $\mathcal A_u$. However, the fact that these sum rules involve a large number of amplitudes renders them phenomenologically somewhat less useful.
Note that the ratios of CKM factors appearing here are approximately invariant under CP transformations.

In the SU(3) limit (S1), both sides of relation~(\ref{eq::SRfull2}) vanish separately, and similarly Eq.~\eqref{eq::SRfull1} can be separated into two parts. The resulting four simpler rules read
\begin{eqnarray}
\mathcal A_{\bar B_s\to D^-D_s^+}+\mathcal A_{\bar B^0\to D_s^-D_s^+}&=& \mathcal A_{\bar B^0\to D^-D^+}\,,\label{eq::rule1}\\
\mathcal A_{\bar B^0\to D_s^-D^+}+\mathcal A_{\bar B_s\to D^-D^+} &=& \mathcal A_{\bar B_s\to D_s^-D_s^+}\,,\label{eq::rule2}
\end{eqnarray}
which can be understood as a consequence of $U$-spin symmetry, see also Ref.~\cite{Gronau:2008ed}, and two more generic SU(3) relations,
\begin{eqnarray}
&&\mathcal A_{\bar B_s\to D^-D^+}+\mathcal A_{\bar B_s\to \bar D^0D^0} \nonumber\\&&\qquad= \frac{\lambda_{us}}{\lambda_{ud}} \left[\mathcal A_{\bar B^0\to D_s^-D_s^+}+\mathcal A_{\bar B^0\to \bar D^0D^0}\right]\,,\\
&&\mathcal A_{\bar B^0\to D_s^-D^+}-\mathcal A_{B^-\to D^-_s D^0}\nonumber\\&&\qquad= \frac{\lambda_{us}}{\lambda_{ud}}\left[\mathcal A_{\bar B_s\to D^-D_s^+}-\mathcal A_{B^-\to D^- D^0}\right]\,.\label{eq::rule4}
\end{eqnarray}
Furthermore, as noted above, $\mathcal A_u(\mathcal D)\leq \mathcal O(\delta^3)$ holds for all modes.

In S2, there are six sum rules which hold up to $\mathcal O(\delta^3)$ or better. They can be chosen as Eqs.~\eqref{eq::isorel1full}-\eqref{eq::isorel3}, Eq.~\eqref{eq::SRfull2}, and two simpler rules\footnote{The  form in which Eq.~\eqref{eq::SR5} is given is such that it is broken only by $\delta A^c_{1}$; it can be simplified to $\mathcal A_{\bar B_s\to D^-D^+}=\lambda_{cs}/\lambda_{cd}\mathcal A_{\bar B^0\to D_s^-D_s^+}$ when both corrections $\delta A^c_{i}$ vanish.}
\begin{eqnarray}
\label{eq::SR4}\mathcal A_{\bar B^0\to D_s^-D_s^+}&=&-\mathcal A_{\bar B^0\to D^0\bar D^0}\,\mbox{ and}\\
\label{eq::SR5}\mathcal A_{\bar B_s\to D^-D^+}&=& \frac{\lambda_{cs}}{2\lambda_{cd}}(\mathcal A_{\bar B^0\to D_s^-D_s^+}-\mathcal A_{\bar B^0\to D^0\bar D^0}),
\end{eqnarray}
both broken by $\delta A^c_i$ and $A^u_i$. Note that these latter rules are broken formally at the same order, but have \emph{relative} corrections of order $\delta$. 
In this scenario $\mathcal A_u(\mathcal D)\leq \mathcal O(\delta^3)$ holds again for all modes.
The number of sum rules can be easily understood from the fact that there are only four MEs up to $\mathcal O(\delta^2)$ and ten amplitudes. 

In S3, actually the same relations hold on the same level, despite the assumption of enhanced penguins (leading to $\mathcal A_u(\mathcal D)\leq \mathcal O(\delta^2)$, only). This fact is related to the RI discussed above and demonstrates that identical sum rules can represent different physical situations.

\subsection{Rate sum rules\label{sec::sumrulesr}}
The fact that in the SU(3) limit the decays have pairwise equal decompositions, see Table~\ref{tab::topamps}, results from the $U$-spin subgroup of SU(3). As a consequence, the well-known equality 
\begin{equation}\label{eq::Uspinpartners1}
\frac{\Gamma(\mathcal{D}_{b\to s})}{\Gamma(\mathcal{D}_{b\to d})}=-\frac{A_{CP}(\mathcal{D}_{b\to d})}{A_{CP}(\mathcal{D}_{b\to s})}
\end{equation}
holds for each pair \cite{FleischerPsiK,Gronau:2000zy}; these relations receive corrections of $\mathcal O(\delta)$ when breaking SU(3).
They are examples of rate sum rules, \emph{i.e.} sum rules formulated directly on the level of observables instead of amplitudes. These are easier to test experimentally; all linear relations between 
\begin{equation}
\Gamma_{\mathcal D}=\frac{\Gamma(\mathcal D)+\Gamma(\bar{\mathcal D})}{2}\mbox{ and } \Delta \Gamma_{\mathcal D}=\frac{\Gamma(\mathcal D)-\Gamma(\bar{\mathcal D})}{2}
\end{equation}
can be found with an algorithm recently discussed in Ref.~\cite{Grossman:2012ry}, which will be applied in the following. From the amplitude sum rules derived above, on the other hand, one can infer typically only inequalities for observables, but they can still be tested in the context of a global analysis.
The quasi-isospin sum rules imply scenario-independent sum rules for the rates:
\begin{eqnarray}
\Gamma_{\bar B^0\to D_s^-D^+} &=& \Gamma_{B^-\to D_s^-D^0}\left(1+\mathcal O(\delta^5)\right)\,,\label{eq::isorelrate1}\\
\Gamma_{\bar B_s\to \bar D^0D^0} &=& \Gamma_{\bar B_s\to D^-D^+}\left(1+\mathcal O(\delta^4)\right)\,,\label{eq::isorelrate2}\\
\Gamma_{\bar B^0\to D^-D^+} &=& \Gamma_{B^-\to D^-D^0}\left(1+\mathcal O(\delta^2)\right)\,.\label{eq::isorelrate3}
\end{eqnarray}
They hold analogously for $B\to DD^*$ and $B\to D^*D^*$ modes. Since there are two sets of modes in $B\to DD^*$, we can define the following double-ratio (analogously, two pairs of transversity amplitudes can be chosen for $B\to D^*D^*$):
\begin{equation}\label{eq::doubleratio}
\frac{BR_{\bar B^0\to D_s^{*-}D^+}}{BR_{B^-\to D_s^{*-}D^0}}\left/\frac{BR_{\bar B^0\to D_s^{-}D^{*+}}}{BR_{B^-\to D_s^{-}D^{*0}}}\right.=1+\mathcal O(\delta^5)\,.
\end{equation}
The size of the corresponding CP asymmetries depends on the scenario, as do other additional sum rules.

For S1, the rate sum rules correspond simply to Eq.~\eqref{eq::Uspinpartners1}; furthermore, $A_{\rm CP}(\mathcal D)\sim \mathcal O(\delta^{5(4)})$ holds for all tree-dominated (annihilation) $b\to s$ modes and $A_{\rm CP}(\mathcal D)\sim \mathcal O(\delta^{3(2)})$ for the $b\to d$ ones. This renders the relations in Eq.~\eqref{eq::Uspinpartners1} very hard to test experimentally. However, they can be replaced by the following relation for the tree-dominated (annihilation) $U$-spin partner modes:
\begin{equation}\label{eq::Uspinpartners2}
\frac{\Gamma(\mathcal{D}_{b\to s})}{\Gamma(\mathcal{D}_{b\to d})}=\left|\frac{\lambda_{cs}}{\lambda_{cd}}\right|^2\left[1+\mathcal O\left(\delta^{3(2)}\right)\right]\,.
\end{equation} 

In S2, the expectations for  the individual CP asymmetries remain the same as in S1,
while Eq.~\eqref{eq::Uspinpartners2} holds only up to $\mathcal O(\delta)$. Since also Eqs.~\eqref{eq::SR4},\eqref{eq::SR5} hold only up to relative order $\mathcal{O}(\delta)$, the same is true for the resulting rate relations:
\begin{eqnarray}
\Gamma_{\bar B^0\to D_s^-D_s^+} &=& \Gamma_{\bar B^0\to D^0\bar D^0}\left(1+\mathcal O(\delta)\right)\,,\\
\left|\frac{\lambda_{cd}}{\lambda_{cs}}\right|^2\Gamma_{\bar B_s\to D^-D^+} &=& \Gamma_{\bar B^0\to D_s^-D_s^+}\left(1+\mathcal O(\delta)\right)\,.
\end{eqnarray}
They are dominantly broken by $\delta A^c_i$; while both are broken by $\delta A^c_{2}$, the first of these relations survives in the presence of $\delta A^c_{1}$. This implies that they can be used to test for the size of these SU(3)-breaking contributions.

In S3, all individual CP asymmetries are expected to be larger, $A_{\rm CP}(\mathcal D)\sim \mathcal O(\delta^{4(3)})$ for tree-dominated (annihilation) $b\to s$ modes and $A_{\rm CP}(\mathcal D)\sim \mathcal O(\delta^{2(1)})$ for $b\to d$ ones. In this case, the relations
\begin{eqnarray}\label{eq::ACPiso}
A_{\rm CP}^{\bar B^0\to D_s^-D^+}     &=& A_{\rm CP}^{B^-\to D^-_sD^0}\left[1+\mathcal O(\delta)\right]\quad{\rm and}\\
A_{\rm CP}^{\bar B_s\to \bar{D}^0D^0} &=& A_{\rm CP}^{\bar B_s\to D^-D^+}\left[1+\mathcal O(\delta)\right]
\end{eqnarray}
hold,
but given the still small absolute size of all $b\to s$ asymmetries, this will again be hard to test.
Finally, there is one non-trivial rule, 
\begin{eqnarray}
&&\Delta\Gamma_{\bar B_s\to D^-D_s^+}+\Delta\Gamma_{\bar B_s\to D_s^-D_s^+}+\Delta\Gamma_{\bar B^0\to D^-D^+}=\nonumber\\
&&\quad \left(2\Delta\Gamma_{B^-\to D^-D^0}+\Delta\Gamma_{B^-\to D_s^-D^0}\right)\left[1+\mathcal O(\delta)\right]\,,
\end{eqnarray}
which can be experimentally verified in the future as another crosscheck of the assumptions in this scenario.

\subsection{Branching ratios and SU(3) breaking\label{sec::BR}}

Eqs.~\eqref{eq::isorelrate1}-\eqref{eq::isorelrate3} provide relations between decay rates, two of which hold to very high precision in the SM. Eq.~\eqref{eq::isorelrate2} is experimentally well fulfilled within the still sizable uncertainties. Using this relation, the measurements can be combined to predict for the SM
\begin{eqnarray}
BR(\bar B_s\to \bar D^0D^0)&=&BR(\bar B_s\to D^-D^+)\nonumber\\&=&(0.21\pm0.03)\times 10^{-3}\,.
\end{eqnarray}
The inequality implied by Eq.~\eqref{eq::isorel3} is fulfilled as well, while Eq.~\eqref{eq::isorelrate3} holds only marginally; this indicates a relatively large annihilation amplitude. More importantly, the data show a tension at the $2\sigma$-level for Eq.~\eqref{eq::isorelrate1}: the corresponding ratio of branching ratios, as measured in Ref.~\cite{Aaij:2013fha}, is expected to decrease to $\sim 1.08$ with additional data, while a confirmation of the present central value with improved precision would challenge the SM: 
\begin{itemize}
\item For the corrections discussed above to yield an effect of this size, the power counting would have to be completely invalidated, for which there are no indications.
\item Isospin-breaking corrections to Eq.~\eqref{eq::isorel1full} could stem from different production rates of neutral and charged $B$ mesons, assumed to be equal in Ref.~\cite{Aaij:2013fha}, or from the decay rates themselves. However, both effects are too small to explain the present central value. Note that the ratio of production rates cancels for the double-ratio defined in Eq.~\eqref{eq::doubleratio}. This double-ratio is consistent with unity with present data~\cite{Beringer:1900zz}, within rather large uncertainties.
\end{itemize}
A confirmation at the level of the present central value would therefore be a sign for a $\Delta I=1$ NP contribution; this possibility is further discussed in Sec.~\ref{sec::NP}.

\begin{figure}
\includegraphics[width=7.4cm]{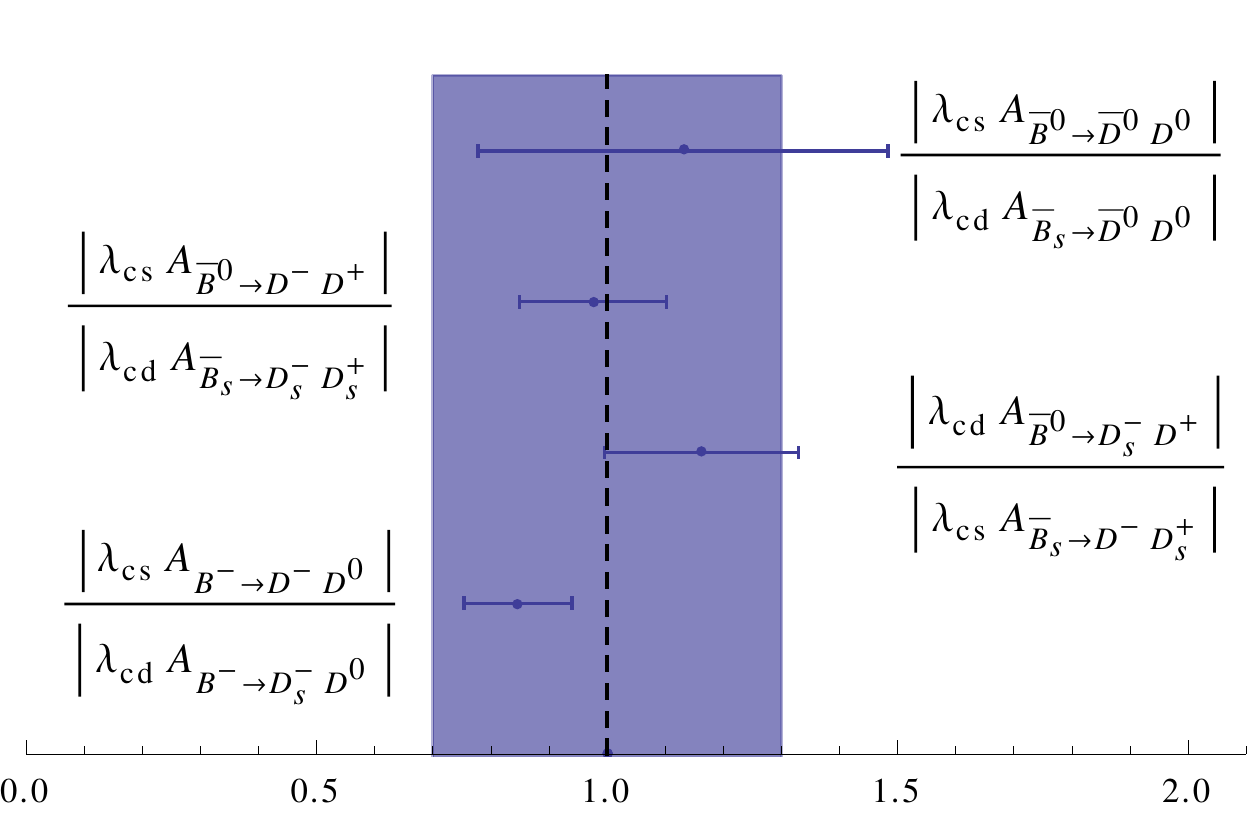}
\caption{\label{fig::Uspin} Experimentally extracted ratios of amplitudes for U-spin partners, cf. Eq.~\eqref{eq::Uspinpartners2}, consistent with the  expected small to moderate SU(3) breaking.}
\end{figure}

To get a first impression of the size of SU(3) breaking in these decays, only the leading amplitudes $T$ and $A^c$ are considered.
In this limit, all direct CP asymmetries vanish and Eq.~\eqref{eq::Uspinpartners1} is replaced by Eq.~\eqref{eq::Uspinpartners2}. The corresponding measurements of ratios of amplitudes are plotted in Fig.~\ref{fig::Uspin}, and indicate small to moderate SU(3) breaking.
This strict limit allows furthermore to extract several values for $T$ (assumed without loss of generality to be real and positive in the following) and $A^c$ from data, separately for the $b\to s$ and $b\to d$ modes. 
Consistent values are extracted for $b\to s$ and $b\to d$ and within each class, showing again no sign of large SU(3) breaking, and also no sign of penguin contributions affecting the rates. The relative size of the two considered amplitudes is
$|A^c/T|\sim15\%$, with a strong phase difference ${\rm arg}(A^c/T)\sim\pi$. While the magnitude is consistent with the expectation from the power counting, $\mathcal{O}(\delta^2)$ -- although on the high side, as already expected from the discussion above -- it is questioning the common neglect of this amplitude in phenomenological analyses. The absolute value is about the size of the estimate in Ref.~\cite{Eeg:2003yq} (albeit the real part of the ratio has the opposite sign), but smaller than the result in Ref.~\cite{Eeg:2005au}. On the other hand the measured values for the branching ratios of the annihilation-dominated modes seem consistently larger than the estimates in Ref.~\cite{Gronau:2012gs}.
The size of $|A^c|$ can actually account for the fact, emphasized in Ref.~\cite{Aaij:2013fha}, that the ratio $BR(\bar B_s\to D_s^+D_s^-)/BR(\bar B^0\to D^+D_s^-)$ shows a significant deviation from the naive expectation of unity holding for $|A^c|\ll |T|$ in the SU(3) limit: using the $U$-spin relation~\eqref{eq::rule2}, a range compatible with the other measurements is obtained for a relative strong phase around $\pi$ between $T$ and $A^c$.

Together these observations indicate the validity of the power counting and a small to moderate SU(3) breaking of about $10-30\%$ on the amplitude level.
While sizable contributions to $\mathcal A_u(\mathcal{D})$ are not excluded, SU(3)-breaking contributions in $\mathcal A_c(\mathcal{D})$ seem to be sufficient to describe the data for branching ratios for the considered modes.
While sizable contributions to $\mathcal A_u(\mathcal{D})$ are not excluded, SU(3)-breaking contributions in $\mathcal A_c(\mathcal{D})$ seem to be sufficient to describe the data for branching ratios for the considered modes.
Note that the apparent anti-alignment of $T$ and $A^c$, chosen here to be along the real axis, implies that ${\rm Re}(\mathcal A_u(\mathcal{D}))$ can be relatively large in all modes without inducing sizable direct CP violation.

\subsection{CP violation\label{sec::CPV}}
CP-violating observables usually constitute the main interest in $B\to DD$ decays. They provide access to fundamental parameters of the SM as well as a high sensitivity to NP. However, data for the branching ratios, apart from being important in their own right, are also necessary to obtain sufficient control over the hadronic uncertainties to extract precision results.

Sensitivity to $\phi_{s}$ stems mainly from time-dependent measurements in $B\to D^+D^-$ and $B_s\to D_s^+D_s^-$; the latter is,
from the theoretical point of view, similarly ``golden'' as $B_s\to J/\psi\phi$. 
Experimentally, however, the $J/\Psi$ mode is advantageous despite the necessity of an angular analysis.
The extraction of the mixing phase $\phi_d$ is difficult, since it enters only in $b\to d$ transitions for $B\to DD$. It has therefore been proposed to use it as an auxiliary channel, 
constraining the shift due to penguin contributions in $S_{\rm CP}(\bar B_s\to D_s^-D_s^+)$ \cite{FleischerPsiK}.
However, this kind of strategy requires additional information to control the influence from SU(3) breaking \cite{Jung:2012mp}; this is achieved within the larger framework developed in this article, 
allowing for model-independently extracting the potential shift in $\phi_s$ due to penguin pollution.

Regarding the CKM angle $\gamma$, the situation is very complicated: first of all, 
obviously significantly measured direct CP asymmetries are required to obtain sensitivity. More importantly, the approximate RI discussed in Sec.~\ref{sec::RI} renders a precision extraction impossible without further theory input. We will therefore use external input for this phase.

The comparison of the extracted value for $\phi_s$ with the one obtained independently yields information on NP.
While the extraction assuming $\mathcal A_u\equiv 0$ 
is trivial once the necessary data are available, this article aims at improving the resulting precision by taking subleading contributions into account. As noted in the introduction, this is of special importance since all measurements so far indicate that NP does not yield large contributions, which therefore compete with the subleading terms in the SM.  Given the possibility discussed in this article to include corrections to the already rather clean SM predictions for various CP asymmetries in these modes, $B\to DD$ decays provide an opportunity for a clean NP search.

The direct CP asymmetries and penguin shifts in $b\to s$ modes are expected to be tiny in the SM for all decay modes: below one (few) per cent in S2 for tree-dominated (annihilation) modes and even with enhanced penguins below few (ten) per cent. 
A significant measurement of a direct CP asymmetry or penguin shift in a $b\to s$ mode outside these ranges would therefore constitute a ``smoking gun'' signal for NP. Additionally, the difference of asymmetries for the quasi-isospin related modes provide direct access to the amplitudes $A_i^u$.

$b\to d$ modes provide additional sensitivity: while they can have a larger SM ``background'', there are strong correlations which again allow to test the SM. Despite the limited available data for CP asymmetries, these correlations provide already information on the modes not measured yet. 

CP~asymmetries in the modes without a tree contribution offer complementary information: due to the suppression of the leading amplitude, they are expected to be larger than their counterparts in tree-dominated modes in the SM; by the same token, the relative influence of NP operators is enhanced. While a single measurement would be insufficient to claim NP, sensitivity is provided by the patterns induced by the flavour symmetry in the SM as well as in given NP models. For example, the SM asymmetries are expected to be correlated due to the SU(3)-limit relation Eq.~\eqref{eq::Uspinpartners1}, softened by SU(3) breaking. For the corresponding $B_s$ modes, the integrated asymmetries are tiny: they are $b\to s$ decays with very small asymmetries to begin with, but are then additionally suppressed by a factor $\sim 1/x_s\sim {\rm few}\%$. 

Direct CP asymmetries and penguin shifts have furthermore the advantage of providing a clear signal for $\mathcal A_u\neq0$; a significant measurement implies a lower bound on $|\mathcal A_u|$ in the corresponding channel.
However, as relative strong phases are involved, the combination of different measurements is necessary before a small $|\mathcal A_u|$ can be deduced from, \emph{e.g.}, a small direct CP asymmetry. 

The direct CP asymmetries for $B^-\to D^0D^-$ and $\bar B^0\to D^+D_s^-$ are measured to be small and consistent with zero, as expected in S2.\footnote{As $A_{\rm CP}(\bar B^0\to D^+D_s^-)$ has been obtained as a crosscheck in the analysis of $B^0\to D^+D^-$ without assessing the systematic uncertainties we do not use the result in the global fit, but see it as another indication for $\mathcal A_u$ being of the expected size. A refined measurement of this mode would be of interest in the context of the present ana\-ly\-sis.} The difficulties regarding the available data for the time-dependent CP asymmetry in $\bar B^0\to D^+D^-$ have already been mentioned in Sec.~\ref{sec::exp}; while the BaBar data are compatible with vanishing $\Delta S_{\rm CP}$ and $A_{\rm CP}$, Belle obtains large central values for both observables, with a significance of around two standard deviations each, but in the unphysical region. The confirmation of large values would imply a violation of the generic SM power counting (S2) which predicts asymmetries in tree-dominated $b\to d$ decays below $10\%$. 
Separate analyses for the two datasets are performed, using either the HFAG average of the two results (a) or the BaBar result only (b), in order to highlight the different predictions.
In any case, new results are expected for this mode as well, hopefully resolving the issue.
Note that, as a remnant of Eq.~\eqref{eq::isorel3}, the leading contributions to $\bar B^0\to D^-D^+$ and $B^-\to D^-D^0$ are equal including their SU(3)-breaking parts, implying the corresponding CP asymmetries to be equal as well up to a few per cent. This implies a SM value closer to the BaBar result. 
More quantitative predictions are made in the context of the following global analysis.

\subsection{Global analysis\label{sec::global}}
The global analysis provides quantitative control over SU(3)-breaking contributions and other subleading terms. The focus lies on the scenarios defined in Sec.~\ref{sec::scenarios}. First their compatibility with the data is analyzed in global $\chi^2$-fits, before presenting the predictions for branching ratios and CP asymmetries. The fits are carried out using the augmented lagrangian \cite{Conn:1991:GCA:107513.107528, Birgin:2008:IUC:1451537.1451539} and Sbplx/Subplex algorithms \cite{Rowan:1990,NLopt} that are implemented in the \lq\lq{}NLopt\rq\rq{} code \cite{NLopt}.

The $\chi^2$ analysis allows for the following observations (see Appendix~\ref{app::detailsfit} for details):
\begin{itemize}
\item S1 generally does not describe the data well.
\item S2 provides an excellent fit, the best of any scenario, since the data are in accordance with SM expectations. 
The fit shows no sign of imaginary contributions to $\mathcal A_c$ and does not improve when including enhanced penguins or symmetry breaking for $A^c$.
\item For dataset (a), only S3 yields a reasonable fit.
\item Dataset (b) is significantly preferred over dataset (a) in the sense that S2 is viable, but the datasets lead to similar results when allowing for arbitrary penguin contributions.
\item In all scenarios the most important amplitude apart from the tree contribution is $A^c$, which is obvious from the significant measurements of annihilation modes in Ref.~\cite{Aaij:2013fha}. The fits confirm the previous observations that $|A^c|$ is on the large side, $|A^c|/T\in [0.1,0.27]$ ($95\%$), and has a negative real part.
\item The fits are consistent with small to moderate ($10-30\%$) SU(3) breaking. The imaginary parts of $\delta T_i$ are only loosely constrained, since they enter observables always doubly suppressed.
\item No indication is found for sizable contributions from $\delta A^c_i$ or $A_i^u$.
\end{itemize}

In the following results from S2 -- corresponding to the SM predictions -- and S3(a/b) are considered in more detail, illustrating the effect of enhanced penguins on the one hand and the different results for the CP asymmetries in $\bar B^0\to D^-D^+$ on the other. In the fit subleading terms such as $\delta A^c_i$ and $A^u_i$ are included, but restricted to lie within the (conservative) ranges expected from their power counting. 
As so far data for CP-violating observables in $B\to DD$ are scarce, the numerical analysis in this section can only be understood as a first step, providing a strategy for the future; specifically, the data do not yet suffice to extract a competitive value for the weak phases $\phi_{s}$, for which therefore external input is used~\cite{CKMfitter}. Nevertheless, the present measurements already yield interesting predictions.  

In order for the relations between observables and MEs to be more transparent, we list in Table~\ref{tab::obscombinations} examples for combinations of observables that are sensitive to a specific parameter to leading order in the power counting. 

For the results regarding branching ratios it is not differentiated between the different scenarios, since the results are almost identical. 
Conservative predictions require the inclusion of the SU(3)-breaking terms also for the annihilation modes: for example, the fit yields the prediction 
\begin{eqnarray}
BR(\bar B^0\to D_s^-D_s^+)&\in&[0.3,2.9]\times10^{-5}\,\,(1\sigma)\,,\\
&\in&[0.1,3.8]\times10^{-5}\,\,(2\sigma)\,,
\end{eqnarray}
while neglecting $\delta A^c_i$ and $\mathcal A_u$ for these modes would yield $BR(\bar B^0\to D_s^-D_s^+)=(1.12\pm 0.15)\times 10^{-5}$, using the experimental values for the other annihilation-dominated modes. 

\begin{figure}[tb]
\begin{centering}
\includegraphics[height=7.cm]{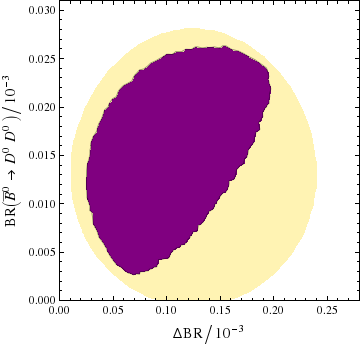}
\end{centering}
\caption{\label{fig::DeltaBRBRB0D0D0} SM fit result for BR($\bar B^0\to \bar D^0D^0$) vs. $\Delta$BR at $95\%$~CL (S2,3a/b together in purple), together with the experimental results (yellow).} 
\end{figure}

The main relations sensitive to NP are Eqs.~\eqref{eq::isorel1full}, \eqref{eq::isorel2full} and to some extent~\eqref{eq::isorel3}. The data confirm these relations, showing however some tension with Eq.~\eqref{eq::isorel1full}, as discussed above. Fig.~\ref{fig::DeltaBRBRB0D0D0} visualizes aspects of Eq.~\eqref{eq::isorel3}. It results in a correlation between the difference of branching ratios
\begin{equation}
\Delta BR \equiv BR(B^-\to D^0 D^-)-r_{\tau,{\rm PS}}BR(\bar B^0\to D^-D^+)\,,
\end{equation}
where $r_{\tau,{\rm PS}}$ denotes the ratio of lifetimes and phase space factors for the two decays, and $BR(\bar B^0\to \bar D^0D^0)$, resulting in both quantities being stronger constrained from the fit than by data directly:
\begin{eqnarray}
\Delta BR &\in&[0.09,0.16]\times10^{-3}\,\,(1\sigma)\,,\\
&\in&[0.05,0.19]\times10^{-3}\,\,(2\sigma)\,,
\end{eqnarray}
to be compared with $\Delta BR|_{\rm exp}=(0.144\pm0.046)\times 10^{-3}$, cf. Table~\ref{tab::data}, and
\begin{equation}
BR(\bar B^0\to \bar D^0D^0)=(1.4\pm0.5)\times 10^{-5}\,.
\end{equation}
The latter result demonstrates again the substantial impact of SU(3) breaking for the annihilation amplitudes, since the SU(3)-limit prediction from data is $BR(\bar B^0\to \bar D^0D^0)=(1.19\pm 0.16)\times 10^{-5}$. 
Therefore all branching ratios of annihilation modes are better constrained in the fit than by their individual measurements available so far; improved data will allow to restrict the SU(3) breaking for these modes, test the quasi-isospin relation Eq.~\eqref{eq::isorel2full} and the correlation from Eq.~\eqref{eq::isorel3}.

\begin{figure}[bt]
\begin{centering}
\includegraphics[height=7cm]{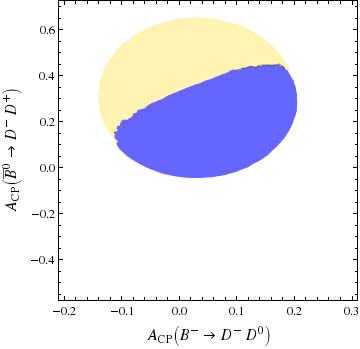}\\[1.2ex]
\includegraphics[height=7cm]{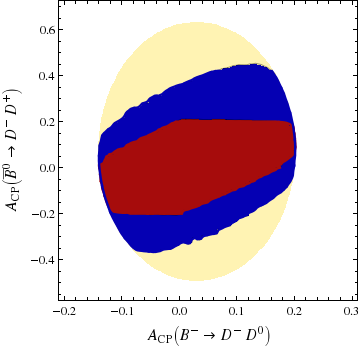}
\end{centering}
\caption{\label{fig::ACPBcDcD0ACPB0DcDc} SM fit result for $A_{CP}(\bar B^0\to D^-D^+)$ vs. $A_{CP}(B^-\to D^-D^0)$ for the datasets (a) (upper plot) and (b) (lower plot), together with the corresponding experimental results. Here and in the following blue areas indicate fits in S3 (light blue for dataset (a) and dark blue for (b)) and red the result in S2.} 
\end{figure}

\begin{figure}
\begin{centering}
\includegraphics[height=7cm]{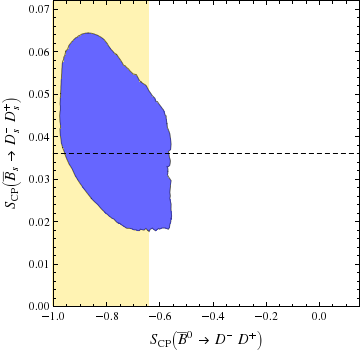}\\[1.2ex]
\includegraphics[height=7cm]{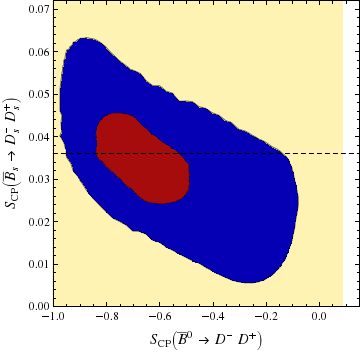}
\end{centering}
\caption{\label{fig::SCPB0DcDcSCPBsDsDs} SM fit result for $S_{CP}(\bar B_s\to D^-_sD^+_s)$ vs. $S_{CP}(\bar B^0\to D^-D^+)$ for the datasets (a) (upper plot) and (b) (lower plot), together with the corresponding experimental results for $S_{CP}(\bar B^0\to D^-D^+)$. Colours as in Fig.~\ref{fig::ACPBcDcD0ACPB0DcDc}.} 
\end{figure}

In Figs.~\ref{fig::ACPBcDcD0ACPB0DcDc} and~\ref{fig::SCPB0DcDcSCPBsDsDs} 
the presently available data for CP asymmetries are shown together with the fit results in the different 
scenarios. 
Note that, as a consequence of Eq.~\eqref{eq::isorel3} and the suppression of $\mathcal A(\bar B^0\to \bar D^0D^0)$, the fit result for $A_{\rm CP}(\bar B^0\to D^-D^+)$ is already nontrivial with present data (note, however, that in S2 the horizontal limitation is due to parameter restrictions, only the diagonal bounds reflect this correlation).
Specifically, the predictions clearly differ for the two datasets, as can be seen in Fig.~\ref{fig::ACPBcDcD0ACPB0DcDc}: even with enhanced penguins, the present central value in dataset (a) is very large. The confirmation of large values with higher precision could again indicate NP.
Furthermore, as can be seen in Fig.~\ref{fig::SCPB0DcDcSCPBsDsDs}, in S3(a) the prediction for $\Delta S_{\rm CP}(\bar B_s\to D^-_sD_s^+)$ is shifted to positive values by the measurement of $S_{\rm CP}(\bar B^0\to D^-D^+)$, to be compared with a distribution around zero for S3(b). The visible correlation stems from the relation 
\begin{equation}
\frac{\Delta S_{\rm CP}(\bar B_s\to D^-_sD_s^+)}{\Delta S_{\rm CP}(\bar B^0\to D^-D^+)}=-\lambda^2 \frac{\cos(\phi_s)}{\cos(\phi_d)}+\mathcal{O}(\delta^3)\,.
\end{equation}
More generally, since the shift due to penguin pollution does not exceed the small value $\sin\phi_s$ in any scenario, all of them predict a very small, positive result for $S_{\rm CP}(\bar B_s\to D_s^-D_s^+)$, the range of which can be further reduced with additional data, specifically a smaller uncertainty in $S_{\rm CP}(\bar B^0\to D^-D^+)$.

\begin{figure}
\begin{centering}
\includegraphics[height=7cm]{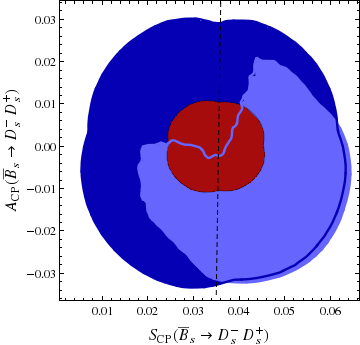}
\end{centering}
\caption{\label{fig::SCPBsDsDsACPBsDsDs} Predictions for  $A_{CP}(\bar B_s\to D^-_sD^+_s)$ vs. $S_{CP}(\bar B_s\to D_s^-D_s^+)$ ($95\%$~CL) from the SM fit for S2 and S3. Colours as in Fig.~\ref{fig::ACPBcDcD0ACPB0DcDc}.} 
\end{figure}

In Fig.~\ref{fig::SCPBsDsDsACPBsDsDs}, the predicted correlation between direct and mixing-induced CP asymmetries in $\bar B_s\to D_s^-D_s^+$ is shown. It is almost absent in S2 and S3(b), while in S3(a) negative values of $\Delta S_{\rm CP}$ tend to imply negative values of the direct CP asymmetry. 

\begin{figure}
\begin{centering}
\includegraphics[height=7cm]{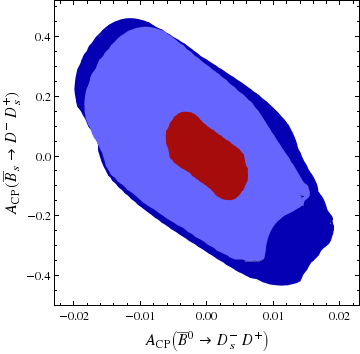}
\end{centering}
\caption{\label{fig::ACPB0DsDcACPBsDcDs} SM predictions for the direct CP asymmetries $A_{CP}(\bar B_s\to D^-D^+_s)$ vs. $A_{CP}(\bar B^0\to D_s^-D^+)$ ($95\%$~CL). Colours as in Fig.~\ref{fig::ACPBcDcD0ACPB0DcDc}.} 
\end{figure}

The correlation for the CP asymmetries of $\bar B^0\to D_s^-D^+$ and $\bar  B_s\to D^-D_s^+$ is shown in Fig.~\ref{fig::ACPB0DsDcACPBsDcDs}. 
Their correlation - stemming from Eq.~\eqref{eq::Uspinpartners1}, but including symmetry-breaking contributions - allows for example to restrict the range of the $b\to s$ mode further by a measurement of the $b\to d$ one or to test the SM when both asymmetries are measured. Note that the large possible range for $A_{\rm CP}(\bar B_s\to D^-D_s^+)$ is not generically expected: it corresponds to the situation in which the penguin amplitude is largely enhanced and additionally the leading amplitude $\mathcal A_c$ reduced compared to $T$ due to SU(3) breaking. Note again that for S3 the CP asymmetry in $B^-\to D_s^-D^0$ is approximately equal to the one in $\bar B^0\to D_s^-D^+$, cf. Eq.~\eqref{eq::ACPiso}.

As mentioned before, the CP asymmetries for the modes without a tree contribution tend to be larger than their tree-dominated counterparts. However, since at present there is no measurement of such  a CP asymmetry available, the global fit does not yield more information than the general correlations already discussed.

\subsection{Prospects for LHCb and Belle~II}
In order to estimate the theoretical uncertainty in the extraction of the weak phase $\phi_s$ from $B\to DD$ in the future, we perform a fit with experimental uncertainties expected in 2022, see Appendix~\ref{app::future}. As is illustrated in Fig.~\ref{fig::DeltaSCP_2022}, the penguin pollution can be well controlled within our approach, so the limiting factor will be the experimental precision for the key observables like $S_{CP}(\bar B_s\to D_s^-D_s^+)$. 

\begin{figure}
\begin{centering}
\includegraphics[height=7cm]{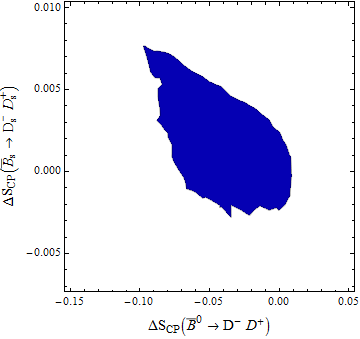}
\end{centering}
\caption{\label{fig::DeltaSCP_2022}Estimated precision for $\Delta S_{\rm CP}(\bar B_s\to D_s^- D_s^+)$ vs. $\Delta S_{\rm CP}(\bar B^0\to D^-D^+)$  ($95\%$~CL) within the future scenario explained in the text, assuming enhanced penguins.} 
\end{figure}

The quasi-isospin sum rules provide an alternative method to search for NP. Here very interesting tests are possible already with the available $3~{\rm fb}^{-1}$ from LHCb and more will be in the farther future. For a potential significant measurement of an intermediate difference of $5$ to $10\%$ in one mode, the issue would be to clearly tell the SM from NP. To that aim additional measurements e.g. of the annihilation-dominated modes and in $B\to DD^*$ will be valuable. The former, since the effect might be enhanced there, because of a smaller normalization. The latter, since not only one of the SM sources of isospin violation cancels in the double-ratio Eq.~\eqref{eq::doubleratio}, but also a potential QCD enhancement could be absent in the second case, while the presence of an additional NP operator should be visible in both sets.

\section{Inclusion of NP\label{sec::NP}}
The inclusion of NP contributions in the symmetry framework is in principle
straight-forward, analogous to the treatment in Ref.~\cite{Hiller:2012xm}:
considering NP models with a specific flavour structure, the corresponding specific
pattern of NP contributions can be included into the analysis. Since in general the
NP contributions have different Dirac structures than the SM ones, they belong to
separate representations, even if their flavour structure is identical.\footnote{In
specific models, the relevant operators may in principle be identical to the SM
ones; in this case, simply the coefficient of the corresponding representations is
to be changed.} Therefore the power counting from above can be kept for the SM
contributions, while additional rules specific to the NP model under consideration
have to be included.

While the coupling strength of one such operator can be absorbed into the unknown
ME(s) related to it, its weak phase and the relative coupling strength of the
SU(3)-related operators have to be specified. Since a third weak phase is not
observable without theory input on the MEs due to RI \cite{Botella:2005ks}, it can
without loss of generality be chosen to be ``distributed'' among the two existing SM
structures, but the relative size of these two contributions and of the
SU(3)-related operators (if present) are model-dependent.

At the moment the inclusion of additional contributions in the fit is
difficult, since the number of measurements does not exceed the number 
of parameters significantly. Furthermore, the SM fits work satisfactorily. 
However, the tension of the data with relation~\eqref{eq::isorel1full} motivates the analysis of possible patterns of isospin-breaking contributions which could become apparent with more precise data. 
Corresponding fits are postponed to future work when more data are available.

We consider generic isospin-changing NP operators in the $\Delta B=1$
Hamiltonian:
\begin{align}
\mathcal{H}^{\mathrm{NP}} = &\Delta_s^{\mathrm{NP}} \left(
(b\bar{u}u\bar{s})^{\mathrm{NP}} - (b\bar{d}d\bar{s})^{\mathrm{NP}} \right)
+\nn\\
                          &\Delta_d^{\mathrm{NP}} \left(
(b\bar{u}u\bar{d})^{\mathrm{NP}} -
(b\bar{d}d\bar{d})^{\mathrm{NP}} \right) \label{eq:np-mod}\\ 
        = &\Delta_s^{\mathrm{NP}} \left(
-\mathbf{\overline{6}}_{1,0,-2/3}^{\mathrm{NP}} -
\mathbf{15}_{1,0,-2/3}^{\mathrm{NP}}   \right) +\nn\\
        &\Delta_d^{\mathrm{NP}} \left(-\frac{1}{2}\sqrt{\frac{3}{2}}
\mathbf{3}_{1/2,-1/2,1/3}^{\mathrm{NP}} + \frac{1}{2} \mathbf{\overline{6}}_{1/2,
-1/2, 1/3}^{\mathrm{NP}} +\right.\nn\\
        &\left.\frac{1}{2\sqrt{6}} \mathbf{15}_{1/2,-1/2,1/3}^{\mathrm{NP}} -
\frac{2}{\sqrt{3}} \mathbf{15}_{3/2,-1/2,1/3}^{\mathrm{NP}} \right) \,.
\label{eq:NPHamilton}
\end{align}
The complex parameters $\Delta_{s,d}^{\mathrm{NP}}$ represent the effective 
NP couplings of the $b\rightarrow s,d$ transitions, respectively.
Note that the Dirac structure in Eq.~(\ref{eq:np-mod}) remains unspecified.

The resulting corrections to the quasi-isospin sum rules Eqs.~\eqref{eq::isorel1full}-\eqref{eq::isorel3} are the same as from $A_{1,2}^u$, with the SM coupling strengths $\lambda_{ud,s}$ replaced by $\Delta_{d,s}^{\mathrm{NP}}$.
Consequently, Eq.~\eqref{eq::SRfull1} is broken by these contributions for $\Delta_{d}^{\mathrm{NP}}/\Delta_{s}^{\mathrm{NP}}\neq \lambda_{ud}/\lambda_{us}$. If the SM MEs $A_{1,2}^u$ are negligible, Eq.~\eqref{eq::SRfull1} can be replaced by the corresponding rule involving $\Delta_{d,s}^{\mathrm{NP}}$. Generically, due to their relation to the same operator, both NP contributions would be expected to be of a similar size. The breaking of Eq.~\eqref{eq::isorel2full} is then expected to be enhanced relative to the one to Eq.~\eqref{eq::isorel1full} approximately as $T/{\rm Re}A^c$. 
Furthermore, due to the same underlying
Hamiltonian, isospin-violating operators in $B\rightarrow DD$ 
would also contribute to various other decays, as $B\to D^*D^{(*)}$, discussed above, but also for example to $B\rightarrow J/\psi K$ decays and other $b\to d,s$ modes, see \emph{e.g.} \cite{Feldmann:2008fb,
Fleischer:2001cw}.
While making a quantitative connection between these modes is again complicated by our limited capability to compute hadronic MEs (and they are not related by flavour symmetry), the combination of different decay modes certainly allows for an improved differentiation between NP and the SM.

\section{Conclusions\label{sec::conclusions}}
Given the absence of clear NP signals in flavour physics so far, searches are confronted with the scenario of NP contributions that are comparable with subleading SM ones. Within the framework developed in this article for $B\to DD$ decays, combining an SU(3) analysis with a power counting for various suppression mechanisms, NP can be differentiated from small SM effects in different ways:
\begin{itemize}
\item The mixing angle $\phi_s$ can be extracted very cleanly with additional data, controlling the penguin pollution from subleading amplitudes from data.
\item Very precise quasi-isospin relations, Eqs.~\eqref{eq::isorel1full},\eqref{eq::isorel2full}, allow for testing isospin-changing NP contributions by measuring ratios of branching ratios. The combination with additional measurements, \emph{e.g.} in $B\to DD^*$, provide improved differentiation from SM contributions.
\item The specific pattern of CP asymmetries and branching ratios implied by the SM lead to various predictions derived in this article that can be tested. 
\end{itemize}
Moreover, $B\to DD$ decays allow to extract different topological amplitudes that are suppressed in the SM, thereby enabling tests and improvements of dynamical calculations and models describing them. 
These theoretical features, combined with the experimental prospects at the LHCb and Belle~II experiments, render them valuable NP probes.

\textbf{Note added:} While finishing this work, a measurement of the time-dependent CP asymmetry in $\bar B_s\to D_s^-D_s^+$ has been published~\cite{Aaij:2014ywt}. The values quoted there for $|\lambda|$ and $\phi_s$ correspond to $S(\bar B_s\to D_s^-D_s^+)=-0.02\pm0.17\pm0.02$ and $A_{\rm CP}(\bar B_s\to D_s^-D_s^+)=-0.09^{+0.20}_{-0.16}\pm0.02$. This measurement is consistent with our prediction in Fig.~\ref{fig::SCPBsDsDsACPBsDsDs}, but does not yet allow for constraining the global fit further. Note that the uncertainties are very close to our estimates in Table~\ref{tab::futuredata}.

\acknowledgments{The authors thank Tim Gershon, Steve Blusk, Conor Fitzpatrick and Thomas Kuhr for useful discussions and Christian Hambrock for support regarding the fits.  The work of M.J. is supported in part by the German Bundesministerium f\"ur Bildung und Forschung (BMBF) and by the DFG cluster of excellence ``Origin and Structure of the Universe''. The work of St.Sch. is supported by the BMBF unter contract no.~05H12VKF.}

\appendix

\section{Experimental inputs\label{sec::AppExp}}
In Table~\ref{tab::adddata}, we provide the experimental results on branching ratios as they are used in the numerical analysis.
\begin{table}
\begin{tabular}{l c}\hline\hline
Observable																	& Value	\\\hline\hline
$BR(B^-\to D^- D^0)$														& $(0.37\pm0.04)\times 10^{-3}\,{}^\S$ \cite{Beringer:1900zz}\\
$BR(B^-\to D^-_sD^0)$														& $(10.0\pm1.7)\times10^{-3}$ \cite{Beringer:1900zz}\\
$\frac{BR(B^-\to D^-_s D^0)}{BR(\bar B^0\to D_s^-D^+)}$						& $1.22\pm0.07$ \cite{Aaij:2013fha}\\
$BR(\bar B^0\to D_s^-D^+)$													& $(7.2\pm0.8)\times 10^{-3}$ \cite{Beringer:1900zz}\\
$\frac{f_s}{f_d}\frac{BR(\bar B_s\to D^-D_s^+)}{BR(\bar B^0\to D_s^-D^+)}$	
& $0.0098\pm0.0010^\dagger$ \cite{Aaij:2014pha}\\
$BR(\bar B^0\to D^-D^+)$													& $(0.226\pm0.023)\times 10^{-3}\,{}^\S$ \cite{Aubert:2006ia,Rohrken:2012ta,Amhis:2012bh}\\
$\frac{f_s}{f_d}\frac{BR(\bar B_s\to D_s^-D_s^+)}{BR(\bar B^0\to D_s^-D^+)}$& $0.143\pm0.009^\dagger$ \cite{Aaij:2013fha}\\
$\frac{BR(\bar B_s\to D_s^-D_s^+)}{BR(\bar B^0\to D_s^-D^+)}$				& $0.56\pm0.11^\ddagger$ \cite{Aaltonen:2012mg}\\
$BR(\bar B_s\to D_s^-D_s^+)$												& $(5.9\pm 1.6)\times 10^{-3}$ \cite{Esen:2012yz}\\
$\frac{f_s}{f_d}\frac{BR(\bar B_s\to D^-D^+)}{BR(\bar B^0\to D^-D^+)}$		& $0.28\pm0.05^\dagger$ \cite{Aaij:2013fha}\\
$\frac{BR(\bar B^0\to \bar D^0D^0)}{BR(B^-\to D_s^-D^0)}$					& $0.0014\pm0.0006$ \cite{Aaij:2013fha}\\
$\frac{f_s}{f_d}\frac{BR(\bar B_s\to \bar D^0D^0)}{BR(B^-\to D_s^-D^0)}$	& $0.0048\pm0.0009^\dagger$ \cite{Aaij:2013fha}\\[1.1ex]\hline\hline
\end{tabular}
\caption{\label{tab::adddata} Experimental results for the branching ratios as used in the numerical analysis; the CP asymmetries are given in Table~\ref{tab::data}. In this table, $f_s/f_d=f_s/f_d|_{\rm LHCb}$. ${}^\S$:~Correction for $\Gamma(\Upsilon(4S)\to B^0\bar B^0)\neq \Gamma(\Upsilon(4S)\to B^+B^-)$ included. ${}^\dagger:$~Value calculated from information in the paper. ${}^\ddagger$:~Calculated using $f_s/f_d|_{\rm Tev}=0.328\pm0.039$ \cite{Amhis:2012bh}. 
}
\end{table}
The available values for CP asymmetries are given in Table~\ref{tab::data}. Ratios are used where they have been measured (avoiding double counting of the uncertainty of the normalization modes) and the correlation induced by $f_s/f_d|_{\rm LHCb}$, which enters most of the recent LHCb results \cite{Aaij:2013fha}, is taken into account explicitly. For Belle and CDF, only one result for a $B_s$ decay enters the analysis, so making the corresponding factor explicit yields no advantage.

The definitions for the coefficients in the time-dependent CP asymmetry, cf. Eq.~\eqref{eq::DefCPA}, are given as
\begin{eqnarray}
A_{\rm CP}(\mathcal D) &=& -\frac{1-|\lambda(\mathcal D)|^2}{1+|\lambda(\mathcal D)|^2}\,,\\
S_{\rm CP}(\mathcal D) &=& \frac{2{\rm Im}\lambda(\mathcal D)}{1+|\lambda(\mathcal D)|^2}\,,\\
A_{\Delta \Gamma}(\mathcal D) &=& \frac{2{\rm Re}\lambda(\mathcal D)}{1+|\lambda(\mathcal D)|^2}, \quad\mbox{and}\\
\lambda(\mathcal D)&=&\eta_{\rm CP}^fe^{-i\phi_D}\frac{\mathcal A(\mathcal D)}{\bar{\mathcal A}(\mathcal D)}\,.
\end{eqnarray}

\section{Details of the SU(3) Analysis\label{app::details}}
\begin{table}
\begin{tabular}{l c c c c c c}
\hline\hline
Decay 						& $C_{8\,3}^c$	& $C_{1\,3}^c$		& $C_{8\,15}^u$				& $C_{8\,\bar6}^u$	 	& $C_{8\,3}^u$			& $C_{1\,3}^u$\\\hline\hline
$B^-\to D^-D^0$				& 1				& 0					& $-\sqrt{\frac{3}{40}}$	& $-\sqrt{\frac{1}{12}}$ & $\sqrt{\frac{3}{8}}$		& 0\\
$B^-\to D^-_sD^0$			& 1				& 0					& $-\sqrt{\frac{3}{40}}$ 	& $-\sqrt{\frac{1}{12}}$ & $\sqrt{\frac{3}{8}}$		& 0\\
$\bar{B}^0\to D^-_sD^+$		& 1				& 0					& $\sqrt{\frac{1}{120}}$	& $\sqrt{\frac{1}{12}}$	 & $\sqrt{\frac{3}{8}}$		& 0\\
$\bar{B}_s\to D^-D_s^+$		& 1				& 0					& $\sqrt{\frac{1}{120}}$	& $\sqrt{\frac{1}{12}}$	 & $\sqrt{\frac{3}{8}}$		& 0\\
$\bar{B}^0\to D^-D^+$		& $\frac{2}{3}$	& $-\frac{1}{3}$	& $\sqrt{\frac{1}{30}}$		& 0						 & $\sqrt{\frac{1}{6}}$		& $-\sqrt{\frac{1}{24}}$\\
$\bar{B}_s\to D_s^-D_s^+$	& $\frac{2}{3}$	& $-\frac{1}{3}$	& $\sqrt{\frac{1}{30}}$		& 0						 & $\sqrt{\frac{1}{6}}$		& $-\sqrt{\frac{1}{24}}$\\
$\bar{B}^0\to D_s^-D_s^+$	& $-\frac{1}{3}$& $-\frac{1}{3}$	& $\sqrt{\frac{1}{120}}$	& $-\sqrt{\frac{1}{12}}$ & $-\sqrt{\frac{1}{24}}$	& $-\sqrt{\frac{1}{24}}$\\
$\bar{B}_s\to D^-D^+$		& $-\frac{1}{3}$& $-\frac{1}{3}$	& $\sqrt{\frac{1}{120}}$	& $-\sqrt{\frac{1}{12}}$ & $-\sqrt{\frac{1}{24}}$	& $-\sqrt{\frac{1}{24}}$\\
$\bar{B}^0\to\bar{D}^0D^0$	& $\frac{1}{3}$	& $\frac{1}{3}$		& $\sqrt{\frac{3}{40}}$		& $-\sqrt{\frac{1}{12}}$ & $\sqrt{\frac{1}{24}}$	& $\sqrt{\frac{1}{24}}$\\
$\bar{B}_s\to\bar{D}^0D^0$	& $\frac{1}{3}$	& $\frac{1}{3}$		& $\sqrt{\frac{3}{40}}$		& $-\sqrt{\frac{1}{12}}$ & $\sqrt{\frac{1}{24}}$	& $\sqrt{\frac{1}{24}}$\\
\hline\hline\end{tabular}
\caption{\label{tab::SU3coeffs} Coefficients $C_{f\,R}^U(\mathcal{D})$ of the SU(3) analysis in the SU(3) limit, cf. Eq.~\eqref{eq::CfRUdef}.}
\end{table}
We provide the list of the coefficients $C_{f\,R}^U$ (cf. Eq.~\eqref{eq::CfRUdef}) in Table~\ref{tab::SU3coeffs}.
The relation to the toplogical amplitudes used in Table~\ref{tab::topamps} is as follows:
\begin{eqnarray}
T &=& \ME{8}{3}_c\,,\\
A^c&=& -\frac{1}{3}\left(\ME{8}{3}_c+\ME{1}{3}_c\right)\,,\\
\tilde P_1 &=& -\sqrt{\frac{1}{120}}\left(\sqrt{45}\ME{8}{3}_u+\sqrt{10}\ME{8}{\bar 6}_u+\right.\nonumber\\
&& \left.\ME{8}{15}_u\right)\,,\\
\tilde P_3 &=&\sqrt{\frac{1}{120}}\left[\sqrt{5}\left(\ME{1}{3}_u+\ME{8}{3}_u\right)+\nonumber\right.\\
&&\left.\sqrt{10}\ME{8}{\bar 6}_u-\ME{8}{15}_u\right]\,,\\
A^u_1 &=& -\sqrt{\frac{1}{15}}\left(\sqrt{5}\ME{8}{\bar6}_u+\sqrt{2}\ME{8}{15}_u\right)\,,\\
A^u_2 &=& \sqrt{\frac{1}{15}}\left(\sqrt{5}\ME{8}{\bar6}_u-\sqrt{2}\ME{8}{15}_u\right)\,.
\end{eqnarray}
Note again that these expressions are equivalent to the SU(3) ones, as long as no assumption is made regarding the relative size of the various amplitudes and all of them are considered as complex. Note furthermore that the absorbed coefficients fulfill $\sum |c_i|^2\leq 1$ ($=1$ only for $T$).
The topological amplitudes are in turn linear combinations of the ones introduced in Ref.~\cite{BurasSilvestrini}; in the SU(3) limit, the translation reads
$T=E_1+P_1$, $A^c=A_2+P_3$, $\tilde P_1=P_1^{\rm GIM}-P_1$, $\tilde P_3=P_3^{\rm GIM}-P_3$, using unitarity, \emph{i.e.} $\lambda_{tD}=-\lambda_{cD}-\lambda_{uD}$. 

The coefficients of the SU(3)-breaking reduced MEs are listed in Table~\ref{tab::SU3coeffseps}. 
\begin{table}
\begin{tabular}{l c c c c}
\hline\hline
Decay 						& $C_{8\,15}^{c,\epsilon}$	& $C_{8\,\bar 6}^{c,\epsilon}$	& $C_{8\,3}^{c,\epsilon}$		& $C_{1\,3}^{c,\epsilon}$	\\\hline\hline
$B^-\to D^-D^0$				& $-\sqrt{\frac{1}{80}}$ 	& $-\sqrt{\frac{1}{8}}$			& $-\frac{1}{4}$ 				& 0 						\\
$B^-\to D^-_sD^0$			& $-\sqrt{\frac{1}{20}}$ 	& 0								& $\frac{1}{2}$					& 0							\\
$\bar{B}^0\to D^-_sD^+$		& $-\sqrt{\frac{1}{20}}$	& 0								& $\frac{1}{2}$					& 0							\\
$\bar{B}_s\to D^-D_s^+$		& $\sqrt{\frac{9}{80}}$		& $\sqrt{\frac{1}{8}}$			& $-\frac{1}{4}$				& 0							 \\
$\bar{B}^0\to D^-D^+$		& $-\sqrt{\frac{1}{20}}$	& 0								& $-\frac{1}{6}$				& $\frac{1}{12}$\\
$\bar{B}_s\to D_s^-D_s^+$	& $\sqrt{\frac{1}{5}}$		& 0						 		& $\frac{1}{3}$					& $-\frac{1}{6}$			\\
$\bar{B}^0\to D_s^-D_s^+$	& $\sqrt{\frac{9}{80}}$		& $-\sqrt{\frac{1}{8}}$			& $\frac{1}{12}$				& $\frac{1}{12}$			  \\
$\bar{B}_s\to D^-D^+$		& $-\sqrt{\frac{1}{20}}$	& 0 							& $-\frac{1}{6}$				& $-\frac{1}{6}$			\\
$\bar{B}^0\to\bar{D}^0D^0$	& $\sqrt{\frac{1}{80}}$		& $-\sqrt{\frac{1}{8}}$			& $-\frac{1}{12}$				& $-\frac{1}{12}$			  \\
$\bar{B}_s\to\bar{D}^0D^0$	& $\sqrt{\frac{1}{20}}$		& 0 							& $\frac{1}{6}$					& $\frac{1}{6}$				 \\
\hline\hline
\end{tabular}
\caption{\label{tab::SU3coeffseps} Coefficients for the $SU(3)$-breaking contributions in $B\to DD$ decays due to the $\mathcal{H}_c$ terms.}
\end{table}
The contributions given in Table~\ref{tab::topampseps} are reordered in terms correcting the $T$ and $A^c$ amplitudes in the topological approach. The translation reads:
\begin{eqnarray}
\delta T_1		&=& \frac{1}{2}\ME{8}{3}_{c,\epsilon}-\frac{1}{\sqrt{20}}\ME{8}{15}_{c,\epsilon}\,,\\
\delta T_2 		&=& \frac{1}{2}\ME{8}{3}_{c,\epsilon}+\frac{1}{\sqrt{20}}\ME{8}{15}_{c,\epsilon}+\nonumber\\
				& & \frac{1}{\sqrt{2}}\ME{8}{\bar 6}_{c,\epsilon}\,,\\
\delta A^c_{1}	&=& \frac{1}{6}\left(\ME{1}{3}_{c,\epsilon}+\ME{8}{3}_{c,\epsilon}\right)+\nonumber\\
				& & \sqrt{\frac{1}{20}}\ME{8}{15}_{c,\epsilon}\,,\\
\delta A^c_{2}	&=& \frac{1}{\sqrt{5}}\ME{8}{15}_{c,\epsilon}-\frac{1}{\sqrt{2}}\ME{8}{\bar 6}_{c,\epsilon}\,.
\end{eqnarray}
While corrections to the tree and annihilation amplitudes can be separated, given the structure of the various decay amplitudes, the choice of the linear combinations '1' and '2' is arbitrary and done in a way to obtain simple expressions for the decay amplitudes.

The corrections to the amplitudes $\mathcal A_u(\mathcal D)$ are obtained analogously to the previous ones. While the calculation yields 13 independent MEs, the rank of the corresponding coefficient matrix in this case is only 9. This does not change when including the leading order contributions, so four of the corrections can in fact be absorbed into the leading order MEs. We give here the coefficients for the reduced system, see Table~\ref{tab::SU3coeffsepsu}, corresponding to physical combinations.
\begin{table}
\begin{tabular}{l c c c c c}
\hline\hline
Decay 						& $C_{1\,3_3}^{u,\epsilon}$	& $C_{8\,3_3}^{u,\epsilon}$	& $C_{8\,\bar 6_3}^{u,\epsilon}$		& $C_{8\,15_3}^{u,\epsilon}$	& $C_{8\,15_{\bar 6}}^{u,\epsilon}$	\\\hline\hline
$B^-\to D^-D^0$				& 0 & $-\sqrt{\frac{3}{128}}$ & $-\frac{\sqrt{3}}{8}$ & $-\sqrt{\frac{3}{640}}$ & $-\sqrt{\frac{1}{960}}$\\
$B^-\to D^-_sD^0$			& 0 & $\sqrt{\frac{3}{32}}$ & 0 & $-\sqrt{\frac{3}{160}}$ & $\frac{1}{\sqrt{60}}$ \\
$\bar{B}^0\to D^-_sD^+$		& 0 & $\sqrt{\frac{3}{32}}$ & 0 & $-\sqrt{\frac{3}{160}}$ & $-\frac{1}{ \sqrt{60}}$ \\
$\bar{B}_s\to D^-D_s^+$		& 0 & $-\sqrt{\frac{3}{128}}$ & $\frac{\sqrt{3}}{8}$ & $\sqrt{\frac{27}{640}}$ &
$\sqrt{\frac{3}{320}}$ \\
$\bar{B}^0\to D^-D^+$		& $\frac{1}{\sqrt{384}}$ & $-\frac{1}{ \sqrt{96}}$ & 0 & $-\sqrt{\frac{3}{160}}$ & $-\frac{1}{\sqrt{240}}$ \\
$\bar{B}_s\to D_s^-D_s^+$	& $-\frac{1}{\sqrt{96}}$ & $\frac{1}{\sqrt{24}}$ & 0 & $\sqrt{\frac{3}{40}}$ & 0 \\
$\bar{B}^0\to D_s^-D_s^+$	& $\frac{1}{\sqrt{384}}$ & $\frac{1}{\sqrt{384}}$ & $-\frac{\sqrt{3}}{8}$ & $\sqrt{\frac{27}{640}}$ & $\sqrt{\frac{3}{320}}$ \\
$\bar{B}_s\to D^-D^+$		& $-\frac{1}{\sqrt{96}}$ & $-\frac{1}{\sqrt{96}}$ & 0 & $-\sqrt{\frac{3}{160}}$ & $-\frac{1}{\sqrt{60}}$ \\
$\bar{B}^0\to\bar{D}^0D^0$	& $-\frac{1}{\sqrt{384}}$ & $-\frac{1}{\sqrt{384}}$ & $-\frac{\sqrt{3}}{8}$ & $\sqrt{\frac{3}{640}}$ &
 $\frac{1}{\sqrt{960}}$ \\
$\bar{B}_s\to\bar{D}^0D^0$	& $\frac{1}{\sqrt{96}}$ & $\frac{1}{\sqrt{96}}$ & 0 & $\sqrt{\frac{3}{160}}$ & $-\frac{1}{ \sqrt{60}}$\\
\hline\hline
\end{tabular}
\caption{\label{tab::SU3coeffsepsu} Coefficients for the $SU(3)$-breaking contributions in $B\to DD$ decays due to the $\mathcal{H}_u$ terms. The labels only  indicate the representation with the largest coefficient in the corresponding linear combination, the additional index indicates the relevant tensor product.}
\end{table}

\begin{table}
\begin{tabular}{l c c}
\hline\hline
Decay 						& $C_{8\,15}^{c,\epsilon^{{}_2}}$& $C_{8\,15}^{u,\epsilon^{{}_2}}$\\\hline\hline
$B^-\to D^-D^0$				& $-\frac{1}{3}$ 			& $\phantom{-}0$\\
$B^-\to D^-_sD^0$			& $\phantom{-}\frac{1}{2}$ 	& $\phantom{-}0$\\
$\bar{B}^0\to D^-_sD^+$		& $\phantom{-}\frac{1}{2}$	& $\phantom{-}1$\\
$\bar{B}_s\to D^-D_s^+$		& $\phantom{-}1$			& $\phantom{-}1$\\
$\bar{B}^0\to D^-D^+$		& $-\frac{2}{3}$			& $-1$\\
$\bar{B}_s\to D_s^-D_s^+$	& $-1$						& $-1$\\
$\bar{B}^0\to D_s^-D_s^+$	& $\phantom{-}1$			& $\phantom{-}1$\\
$\bar{B}_s\to D^-D^+$		& $\phantom{-}\frac{1}{2}$	& $\phantom{-}1$\\
$\bar{B}^0\to\bar{D}^0D^0$	& $\phantom{-}\frac{1}{3}$	& $\phantom{-}0$\\
$\bar{B}_s\to\bar{D}^0D^0$	& $-\frac{1}{2}$			& $\phantom{-}0$\\
\hline\hline
\end{tabular}
\caption{\label{tab::SU3coeffseps2} Coefficients for higher-order $SU(3)$-breaking contributions in $B\to DD$ decays (normalized to the largest coefficient).}
\end{table}

\section{\label{app::detailsfit} Fit details}
We provide additional information regarding the fits in Sec.~\ref{sec::global}. First we observe a very good consistency of the available measurements (apart from the CP asymmetries in $\bar B^0\to D^-D^+$). The only sizable offset in $\chi^2$  in the following fits (in the sense that the value cannot be reduced by any parameter choice for realistic values of the MEs)  stems from the tension with the quasi-isospin relation Eq.~\eqref{eq::isorel1full}. To test the overall consistency of a scenario we start with $T$ and $A^c$ and add single MEs (while in the numerical analysis all MEs are always included).

In S1, neglecting SU(3)-breaking contributions, no good fit is achieved. The only acceptable fit with $\chi^2\approx 17$ for 10 degrees of freedom (dof)\footnote{Note that the determination of this number is non-trivial since for example no observable is sensitive to ${\rm Im}\delta T_{1,2}$ to leading order and potential constraints from the power counting. In these fits we only demand $|X/T|\leq 1$ ($X$ being real- or imaginary part of an ME) and count simply observables and parameters to determine the dof.} is reached for dataset (b) when adding only $\tilde P_1$. However, this ME is then required to be very large (larger than in S3), since it is ``used'' to fit some of the rate differences. Restricting $\tilde P_1$ to the expected range again yields a bad fit. For dataset (a), no acceptable fit is found.

The fits for dataset (a) in S2 remain bad, while for dataset (b) an excellent fit with $\chi^2_{\rm min}\approx 9$ for 8~dof exists when adding only $\delta T_{1,2}$ to the fit, as expected with our power counting. Note that neither imaginary parts in $\mathcal A_c$ nor penguin contributions are necessary in this case. Also, the fit does not worsen at all when restricting to the expected SU(3) breaking. 

In S3, at least $\tilde P_1$ should be included, since it is of the same order as $A^c$ in this scenario. That leads to an acceptable fit for dataset (a) with $\chi^2_{\rm min}\simeq11$ for 6 dof. For dataset (b), the fit does not improve significantly compared to S2, and is in that sense worse than the previous fit since there are less dof. The global $\chi^2_{\rm min}$-values when including all MEs with their power counting are $\chi^2_{\rm min, S3(a)}=9.3$ and $\chi^2_{\rm min, S3(b)}=8.9$.

In order to better understand the consequences of certain measurements for the global fit, we present in Table~\ref{tab::obscombinations} examples for combinations of observables that are to leading order sensitive to one parameter in the different scenarios. Listed are the coefficients that multiply the corresponding rates or CP asymmetries, yielding for example for the next-to-last column
\begin{equation}
\frac{A_{CP}(\bar B^0\to D_s^-D^+)-A_{CP}(B^-\to D_s^-D^0)}{2\,{\rm Im}(\lambda_{us}/\lambda_{cs})}=\frac{{\rm Im}(A_1^u)}{T}\,.
\end{equation}

\begin{table*}
\begin{tabular}{l c c c c c c|c c c c}
\hline\hline
Mode						& \multicolumn{6}{c}{$|\mathcal{A_D}/\lambda_{cD}|^2$}	&\multicolumn{4}{|c}{$A_{\rm CP}(\mathcal D)/{\rm Im}(\lambda_{uD}/\lambda_{cD})$}\\\hline\hline
$B^-\to D^- D^0$			& 0				& $\phantom{-}\frac{1}{6}$	& $\phantom{-}0$			& 0					& $\phantom{-}0$			& $\phantom{-}0$				& 0							& 0				& $\phantom{-}0$			& $\phantom{-}0$\\
$B^-\to D^-_s D^0$			& 0				& $\phantom{-}0$			& $\phantom{-}0$			& 0					& $\phantom{-}0$			& $\phantom{-}0$				& 0							& 0				& $-\frac{1}{2}$			& $\phantom{-}0$\\
$\bar B^0\to D_s^-D^+$		& 0				& $\phantom{-}0$			& $\phantom{-}0$			& 0					& $\phantom{-}0$			& $\phantom{-}0$				& $\frac{1}{2}$				& 0				& $\phantom{-}\frac{1}{2}$	& $\phantom{-}0$\\
$\bar B_s\to D^-D_s^+$		& 0				& $-\frac{1}{2}$			& $\phantom{-}0$			& 0					& $\phantom{-}0$			& $\phantom{-}0$				& 0							& 0				& $\phantom{-}0$			& $\phantom{-}0$\\
$\bar B^0\to D^-D^+$		& $\frac{2}{3}$	& $\phantom{-}0$			& $-\frac{1}{3}$			& 0					& $\phantom{-}0$			& $\phantom{-}0$				& 0							& 0				& $\phantom{-}0$			& $\phantom{-}0$\\
$\bar B_s\to D_s^-D_s^+$	& $\frac{1}{3}$	& $\phantom{-}\frac{1}{3}$	& $\phantom{-}\frac{1}{3}$	& 0					& $\phantom{-}0$			& $\phantom{-}0$				& 0							& 0				& $\phantom{-}0$			& $\phantom{-}0$\\
$\bar B^0\to D_s^-D_s^+$	& 0				& $\phantom{-}0$			& $\phantom{-}0$			& $\frac{1}{3}$		& $\phantom{-}\frac{1}{6}$	& $\phantom{-}\frac{1}{2}$		& 0							& 0				& $\phantom{-}0$			& $\phantom{-}0$\\
$\bar B_s\to D^-D^+$		& 0				& $\phantom{-}0$			& $\phantom{-}0$			& $\frac{1}{3}$		& $\phantom{-}0$			& $\phantom{-}0$				& 0							& $\frac{1}{2}$	& $\phantom{-}0$			& $\phantom{-}\frac{1}{2}$\\
$\bar B^0\to \bar D^0D^0$	& 0				& $\phantom{-}0$			& $\phantom{-}0$			& $\frac{1}{3}$		& $\phantom{-}\frac{1}{6}$	& $-\frac{1}{2}$				& 0							& 0				& $\phantom{-}0$			& $\phantom{-}0$\\
$\bar B_s\to \bar D^0D^0$	& 0				& $\phantom{-}0$			& $\phantom{-}0$			& 0					& $-\frac{1}{3}$ 			& $\phantom{-}0$				& 0							& 0				& $\phantom{-}0$			& $-\frac{1}{2}$\\\hline
Par. comb.					& $T^2$			& $T\,{\rm Re}(\delta T_1)$	& $T\,{\rm Re}(\delta T_2)$	& $|A^c|^2$			& ${\rm Re}[(A^c)^*\delta A^c_1]$	  & ${\rm Re}[(A^c)^*\delta A^c_2]$ & ${\rm Im}(\tilde P_1)/T$  & ${\rm Im}(\tilde P_3/A^c)$	& ${\rm Im}(A_1^u)/T$			  & ${\rm Im}(A_2^u/A^c)$\\\hline
Correction					& $\delta^2$	& $\delta^2$				& $\delta^2$				& $\delta^{6(5)}$	& $\delta^{6(5,\dagger)}$				& $\delta^{6}$				& $\delta^{4(3)}$			& $\delta^{3(2)}$				  & $\delta^{4}$ 			  & $\delta^3$\\\hline\hline
\end{tabular}
\caption{\label{tab::obscombinations} Examples for combinations of observables that determine simple fit parameter combinations to leading order, as further explained in the text. The last line indicates the expected absolute size of the corrections to these relations; the number in brackets, when present, shows a possible enhancement of the correction for enhanced penguins. ${}^\dagger$: Note that the enhanced penguin constitutes a \emph{relative} correction of $\mathcal O(1)$ in this case.}
\end{table*}

\section{Future scenarios\label{app::future}}
In order to analyze the future potential of the method developed in this article, we
give projections for the experimental 
uncertainties of the various observables, considering the plans for LHCb and
Belle~II. While we expect these estimates to be 
rather reliable for the statistical uncertainties, the systematic ones are clearly
more complicated; we try to include the expected 
main improvements while still keeping the estimates conservative.

LHCb has recorded another approximate $2~{\rm fb}^{-1}$ at 8~TeV in 2012 which are
analyzed at the moment. The plan is~\cite{Gershon:1604468} 
to record another $5-6~{\rm fb}^{-1}$ at 13~TeV 2015-2018 and after the upgrade
$15~{\rm fb}^{-1}$ at 14~TeV 2020-2022; from 2024 on additional data up to at least 
$50~{\rm fb}^{-1}$ are to be recorded. Belle~II is expecting to start physics runs
in 2018;\footnote{A possible shift of this date by one year or more should be taken into account for the following dates accordingly.} in the first year, one might expect 4 times the Belle 
dataset while by the end of 2022 the aim is to have reached $50~{\rm ab}^{-1}$. 

The resulting yields scale with these expected integrated luminosities, but other
factors have to be taken into account.
In the case of LHCb we additionally account for the higher cross sections for $B$
mesons at $13/14$~TeV by a factor of two, and 
for the improvements from the upgrade by another factor of two. Compared to the
yields in Ref.~\cite{Aaij:2013fha} at $1~{\rm fb}^{-1}$, this 
procedure yields the approximate factors $4$ for the dataset so far,\footnote{We use a factor of $4$ instead of $3$ to account for analysis improvements and the already larger $B$ cross section at $8$~TeV.} 14 until 2018,
and 74 until 2022. Clearly, these factors have large uncertainties. Especially it should be noted that the analyses at higher beam energy are challenging; some uncertainties might also increase in this environment, so the advantages considered here might be partially reversed.

Of the 1000 ${\rm fb}^{-1}$ collected at Belle about 121~fb$^{-1}$ were recorded at
the Y(5S) resonance, and 711 fb$^{-1}$ at the Y(4S) resonance
\cite{Sibidanov:2014wja}.
In our projections we assume that this ratio will remain the same at Belle II. The
expected integrated luminosity at Belle II is 4~ab$^{-1}$ in 2018 \cite{BPAC14:Hara}
and 50~ab$^{-1}$ 
until 2022-2023 \cite{Mohanty:2013kza}.
Consequently, we scale the yields with a factor of four and 50 for 2018 and 2022,
respectively.
From these prospects for the yields we estimate the expected statistical uncertainties.

The systematic uncertainties obviously depend on the observable under consideration. 
For the branching ratios of $B_s$ modes, the limiting systematic uncertainties will
be the external inputs; $f_s/f_d$ and (in part related to that) the 
measurements of the corresponding $D$ decay modes. Especially the former input is already
clearly dominated by systematic/theory uncertainties, see 
Refs.~\cite{Aaij:2013qqa,LHCb-CONF-2013-011}, making it hard to reduce the related
uncertainty below $5\%$. The latter uncertainty is expected to improve with more statistics.

The remaining (ratios of) branching ratios are expected to be determined with higher precision,
although still the $D$ branching ratios enter. For high-statistics measurements one should furthermore include potential differences between $f_d$ and
$f_u$, which are typically assumed to be equal.

For the CP asymmetries in flavour specific modes, systematic uncertainties are
determined by the production and detection asymmetries.
The production asymmetry can in principle be determined from a pure tree decay like
$B^-\to D^0\pi^-$ and also the uncertainty for the detection 
asymmetry is reducible with more statistics. 
This situation implies that the uncertainties for these CP asymmetries 
will probably remain determined by the statistical ones. We scale the uncertainties according to our prescription above where such measurements are available. In the cases where only branching ratios have been measured so far, we estimate the direct CP asymmetry uncertainty from that of the branching ratio. For the LHCb measurements where the available yields are from 
branching ratio measurements,
we add a factor of two in selection efficiency 
for the CP asymmetry measurements
(which is again a rough estimate).
We assume that the uncertainty for $A_{CP}(\bar{B}^0\rightarrow D_s^-D^+)$ is the same as for $A_{CP}(B^-\rightarrow D_s^- D^0)$.  

For time-dependent asymmetries, the issue is more complicated.  The CP-violating 
parameters of interest are obtained from fits, rendering the relation to the number 
of events non-trivial. To estimate this, we use existing time-dependent analyses from LHCb  
for CP asymmetries in $B_s\to K^+K^-$ and $B^0\to \pi^+\pi^-$ decays \cite{Aaij:2013tna}. 
We scale the resulting statistical precision obtained in this analysis with the 
square-root of yields, taking into account the effective tagging power, which we assume to stay at 
$\varepsilon_{\mathrm{eff}} =  5\%$.
In this way we obtain the LHCb
prospects for the time-dependent CP-asymmetries of 
$\bar{B}_s\rightarrow D_s^- D_s^+$,  $\bar{B}_s\rightarrow D^-D^+$,
$\bar{B}^0\rightarrow \bar{D}^0 D^0$ and $\bar{B}_s\rightarrow \bar{D}^0 D^0$. 
For the channel $\bar{B}^0\rightarrow D^- D^+$ we scale the Belle results. 
For $\bar{B}^0\rightarrow D_s^- D_s^+$ 
and $\bar{B}^0\rightarrow \bar{D}^0D^0$ no statistical significant yield is
available from Belle yet.
In order to obtain an estimate for $BR(\bar{B}^0\rightarrow D_s^- D_s^+)$
we scale  the relative error of $\bar{B}^0\rightarrow D^-D^+$ 
to account for the
differences in yields and multiply the result with the corresponding SU(3)-limit 
 branching ratio.
We do not estimate the corresponding time-dependent CP-asymmetry parameters.

These considerations yield the uncertainty estimates given in Table~\ref{tab::futuredata}.

\begin{table}
\begin{tabular}{l c c c c}
\hline\hline
Mode                         & Observable    & $\delta\mathcal O$ (2014)     &
$\delta\mathcal O$ (2018)     & $\delta\mathcal O$ (2022)\\\hline\hline
$B^-\to D^- D^0$            & BR$/10^{-3}$     &   & $0.01$ & $0.004$ \\
                             & $A_{CP}/\%$ &   & $3$ & $1$ \\
%
$B^-\to D^-_s D^0$          & BR$/10^{-3}$       & $0.6$ & $0.2$ & $0.07$ \\
                             & $A_{CP}/\%$ & $0.8$ & $0.4$ & $0.2$ \\
%
$\bar B^0\to D_s^-D^+$     & BR$/10^{-3}$       &   & $0.2$ & $0.06$ \\
                             & $A_{CP}/\%$ & 0.8 & $0.4$ & $0.2$ \\
%
$\bar B_s\to D^-D_s^+$     & BR${}_{\rm rel}$
&   & $0.002$ & $0.0009$ \\
                             & $A_{CP}/\%$ & $5$ & $3$ & $1$ \\
%
$\bar B^0\to D^-D^+$       & BR$/10^{-3}$       &   & $0.008$ & $0.002$ \\
                             & $S,C/\%$    &   & $7$ & $2$ \\
%
$\bar B_s\to D_s^-D_s^+$   & BR${}_{\rm rel}$
     & $0.02$ & $0.008$ & $0.003$ \\
                            & $S,C/\%$    & $16$ & $8$ & $4$ \\
$\bar B^0\to D_s^-D_s^+$   & BR$/10^{-3}$       &   & $ 0.002$ & $ 0.0005$ \\
                             & $S,C/\%$    &    &    & $  $ \\
$\bar B_s\to D^-D^+$       & BR${}_{\rm rel}$
    & $0.1$
& $0.05$ & $0.02$ \\
                             & $S,C/\%$    &  & $27$ & $12$ \\
$\bar B^0\to \bar D^0D^0$  & BR$/10^{-3}$      & $0.003$ & $0.002$ & $0.0007$ \\
                             & $S,C/\%$    &  &  & $23$ \\
$\bar B_s\to \bar D^0D^0$  & BR${}_{\rm rel}$
& $0.002$
& $0.0008$ & $0.0003$ \\
                             & $S,C/\%$    &  & $26$ & $11$ \\\hline\hline
\end{tabular}
\caption{Extrapolated uncertainties for $B\to DD$ observables, assuming future data as described in the text. The index 'rel' indicates that the uncertainty is given for the ratio with the normalization mode in Ref.~\cite{Aaij:2013fha}, \emph{not} the branching ratio itself. For the CP asymmetries a value is only given when our estimate lies below $40\%$ (and the present uncertainties). \label{tab::futuredata}}
\end{table}

\bibliography{BtoDD}

\end{document}